\documentclass[longauth]{aa} 

\usepackage{graphicx}
\usepackage{array,multirow,makecell}
\usepackage{txfonts}
\usepackage[dvipsnames]{xcolor}
\usepackage{hyperref}
\hypersetup{colorlinks,linkcolor={blue},citecolor={blue},urlcolor={red}} 
\usepackage[capitalise]{cleveref}
\usepackage{natbib}
\usepackage{amsmath}
\usepackage{ulem} 
\usepackage{comment}
\bibpunct{(}{)}{;}{a}{}{,} 
\usepackage[normal]{threeparttable}
\usepackage{multicol}
\usepackage{tabularx}
\usepackage{colortbl}
\usepackage[caption = false]{subfig}
\usepackage{float}  
\usepackage{placeins}
\usepackage{lscape} 
\usepackage{pdflscape} 
\usepackage{arydshln} 
\usepackage{booktabs} 
\newcommand{\rom}[1]{\uppercase\expandafter{\romannumeral #1\relax}}
\usepackage{pifont}
\usepackage{nccmath} 
\usepackage{gensymb}
\usepackage{soul}
\usepackage{threeparttablex}

\newcommand{\msun}{\mbox{$M_\odot$}}

\newcommand{\hii}{H\mbox{\sc ~ii} }

\newcommand{\getsf}{\textsl{getsf}\xspace}
\newcommand{\FAMILY}{\textsl{FAMILY}\xspace}

\begin{document} 

\title{ALMA-IMF}
\subtitle{XXII. Role of core subfragmentation in the IMF origin: Hierarchical mass cascade in W43-MM1}
\titlerunning{ALMA-IMF XXII. Role of core subfragmentation in the IMF origin: Hierarchical mass cascade in W43-MM1}

\author{  F.\ Motte \inst{1} 
\and
          N.\ Le Nestour \inst{2, 1} \and
          R.\ Veyry \inst{1} 
\and
          N.\ Brouillet \inst{3} \and
          T.\ Nony \inst{3} \and
          B.\ Thomasson \inst{1} \and
          F.\ Louvet \inst{1} \and
          I.\ Joncour \inst{1} \and
          E.\ Moraux \inst{1} \and
          A.\ Men’shchikov \inst{4} \and
          T.\ Yoo \inst{5} \and
          A.\ Ginsburg \inst{5} \and
          A.\ Gusdorf \inst{6, 7} \and
          A.\ M.\ Stutz \inst{8} \and 
          R.\ Galv\'an-Madrid \inst{9} \and
          T.\ Csengeri \inst{3} \and
          R.\ H.\ \'Alvarez-Guti\'errez \inst{10} \and
          M.\ Armante \inst{11} \and
          Y.\ Bernard \inst{1} \and
          M. Bonfand \inst{12} \and
          S.\ Chevalier \inst{1} \and
          N.\ Cunningham \inst{13} \and
          P.\ Dell'Ova \inst{14} \and          
          M.\ Gonz\'alez \inst{15} \and
          A.\ Koley \inst{8} \and 
          F.\ A.\ Olguin \inst{16, 17} \and
          D.\ Panda \inst{3} \and
          Y.\ Pouteau \inst{18} \and
          J.\ Salinas \inst{9} \and
          P.\ Sanhueza \inst{19} \and
          N.\ A.\ Sandoval-Garrido \inst{8} \and
          M.\ Valeille-Manet \inst{20, 1, 3}
          }

\institute{Univ. Grenoble Alpes, CNRS, IPAG, 38000 Grenoble, France             
    \and LPC2E, CNRS, Univ. Orl\'eans, 45071, Orl\'eans Cedex 02, France    
    \and Laboratoire d'astrophysique de Bordeaux, Univ. Bordeaux, CNRS, B18N, all\'ee Geoffroy Saint-Hilaire, 33615 Pessac, France 
    \and Universit{\'e} Paris-Saclay, Universit{\'e} Paris Cit{\'e}, CEA, CNRS, AIM, 91191 Gif-sur-Yvette, France 
    \and Department of Astronomy, University of Florida, PO Box 112055, USA 
    \and Laboratoire de Physique de l'\'Ecole Normale Sup\'erieure, ENS, Universit\'e PSL, CNRS, Sorbonne Universit\'e, Universit\'e de Paris, Paris, France  
    \and Observatoire de Paris, PSL University, Sorbonne Universit\'e, LUX, 75014, Paris, France    
    \and Departamento de Astronom\'{i}a, Universidad de Concepci\'{o}n, Casilla 160-C, Concepci\'{o}n, Chile 
    \and Instituto de Radioastronom\'ia y Astrof\'isica, Universidad Nacional Aut\'onoma de M\'exico, Morelia, Michoac\'an 58089, M\'exico      
    \and SUPA, School of Physics and Astronomy, University of St. Andrews, North Haugh, St. Andrews KY16 9SS, UK 
    \and INAF—Osservatorio Astrofisico di Arcetri, Largo Enrico Fermi 5, 50125 Firenze, Italy 
    \and Departments of Astronomy and Chemistry, University of Virginia, Charlottesville, VA 22904, USA 
    \and SKA Observatory, Jodrell Bank, Lower Withington, Macclesfield, SK11 9FT, United Kingdom 
    \and Universit\'e Paris-Saclay, CNRS, Institut d’Astrophysique Spatiale, 91405 Orsay, France 
    \and Universidad Internacional de Valencia (VIU), C/Pintor Sorolla 21, E-46002 Valencia, Spain 
    \and Center for Gravitational Physics, Yukawa Institute for Theoretical Physics, Kyoto University, Kitashirakawa Oiwakecho, Sakyo-ku, Kyoto 606-8502, Japan 
    \and National Astronomical Observatory of Japan, National Institutes of Natural Sciences, 2-21-1 Osawa, Mitaka, Tokyo 181-8588, Japan 
    \and Observatoire astronomique des Makes, Saint-Louis 97421, La Réunion, France 
    \and Department of Astronomy, School of Science, The University of Tokyo, 7-3-1, Hongo, Bunkyo-ku,
Tokyo 113-0033, Japan 
    \and Cardiff University, School of Physics \& Astronomy, Queen’s buildings, The parade, Cardiff CF24 3AA, UK 
}

\date{Received April 1, 2026; accepted June 29, 2026}
 
\abstract
{}
%
{The gravo-turbulent fragmentation of the interstellar medium is expected to create a hierarchical cascade of cloud structures, crossing the scales from core to disk. We aim to predict how the currently observed top-heavy core mass function (CMF) in the massive protocluster W43-MM1 evolves due to core subfragmentation.}
%
{We used the \getsf algorithm to extract sources in five ALMA images of W43-MM1 at 3~mm, with a spatial resolution ranging from 14~kau to 270~au. Then, we applied \textsl{FAMILY}, a graph-theory-based analysis tool, to create and characterize networks of nested sources in W43-MM1. We compared the hierarchical fragmentation cascade of W43-MM1 to those measured in the NGC~2264 protocluster and in synthetic images of an Orion-like protocluster simulated by magneto-hydrodynamical calculations.}
%
{Assuming self-similarity, we measure a small fractality index of $\mathcal{F}_{3\rm D}=1.19\pm 0.10$ in W43-MM1, which means that, on average, a cloud structure will fragment into only 1.19 fragments each time the physical scale decreases by a factor of two. In line with values measured above the core scale in the NGC~2264 and synthetic protoclusters, the W43-MM1 fractality index increases by about $\sim$30\% at larger scales. We also estimate an imbalanced mass partition between siblings, with $\frac{2}{3}$ of the mass of siblings at a given scale belonging to the dominant sibling. The mass transfer efficiency, computed from one physical scale to another, is high and corresponds to a core formation efficiency (CFE) from 2400~au cores to 200~au seeds of $\sim$16\%.} 
%
{Based on the fractality and efficiency values measured in W43-MM1, the gravo-turbulent model by Thomasson predicts that its fragmentation below $\sim$14~kau is not driven by turbulence but by gravity. Using these parameters and the measured mass partition, we demonstrate that the seed mass function, from which the the initial mass function (IMF) emerges, has a high-mass end which remains top-heavy. Therefore, based on our current assumptions, core subfragmentation in W43-MM1, and perhaps more broadly in massive Galactic protoclusters, plays a minimal role in shaping the high-mass slope of the IMF. }

\keywords{stars: formation -- stars: IMF -- stars: massive -- ISM: clouds -- submillimeter: ISM -- ISM: dust}

\maketitle

\section{Introduction}
\label{s:intro}

Understanding star formation requires addressing the challenge of determining how the discrete distribution of stars emerges from the continuous distribution of gas in molecular clouds. Most analytical models that have been proposed to describe star formation are based on the interplay between supersonic turbulence, magnetic field, and gravity; they are called gravo-turbulent models \citep[e.g.,][]{hennebelleChabrier2008, hopkins2012}. In line with the fact that gravity and incompressible turbulence are scale-free processes, studies of the interstellar medium (whether atomic, ionized, or molecular) first showed that it exhibits fractal properties \citep[e.g.,][]{elmegreen2001fract, hennebelle2012}. 
Subsequent studies have revealed that the molecular medium is better described by a superposition of several hierarchies of structures that coexist at all scales, and therefore a multifractal model \citep{falgarone2004, miville2007, robitaille2020}. This multifractality can be explained by the emergence of physical processes associated with star formation, as evidenced by the appearance of a network of dense, filamentary structures coherent over a range of physical scales down to the smallest observed scale \citep{elia2018, robitaille2019}. These processes include turbulence intermittency, ambipolar diffusion, conservation of angular momentum, changes in the equation of state, and stellar feedback, such as heating and compression \citep[e.g.,][]{girart2013, robitaille2020, thomasson2024}. The fragmentation cascade associated with these multifractal properties could therefore change, at small scales in the densest cloud areas where optical depth increases, and rotation and magnetic fields intensify \citep[e.g.,][]{thomasson2024}. It could also change close to new-born stars, which heat their surrounding gas, drive outflows and potentially \hii regions. The multifractal characteristics of molecular clouds have only recently begun to be investigated \citep[e.g.,][]{elia2018, pokhrel2018, robitaille2019, thomasson2024}, and further studies are needed to understand how these characteristics vary during the star formation process and within star-forming clouds.

In the fragmentation cascade associated with the multifractal structure of clouds, the gas mass reservoirs used to form individual stars or close binaries are generally called cores \citep[e.g.,][]{mckeeOstriker2007}. Their exact definition is the subject of heated debate, particularly regarding the formation of high-mass stars and whithin synthetic protoclusters \citep[e.g.,][]{smith2009, motte2018a, louvet2021simu, pelkonen2021}. Observationally, cores are identified as dense, gravitationally bound cloud structures with sizes ranging from $\sim$10~kau ($\sim$0.05~pc) to $\sim$2~kau ($\sim$0.01~pc) depending on the gas density of the protocluster and thus its ability to form high-mass stars \citep[e.g.,][]{andre2000, motte2018b}. In nearby, low-mass star-forming regions initially studied, the mass distribution of cores (the core mass function, or CMF) resembles the initial mass function (IMF) of stars \citep[e.g.,][]{motte1998, enoch2008, konyves2015}. Since the shape of the CMF was not observed to vary much, the community considered that the CMF followed the canonical shape of the IMF or, at least, its high-mass ($\sim$1--10~$\msun$) end \citep[e.g.,][]{offner2014, hennebelleGrudic2024}. This led to the interpretation that cores were the direct progenitors of stars \citep[e.g.,][]{andre2000}. However the putative direct link between cores and stars is now being questioned, in particular in dynamical protoclusters. Two main processes have been identified that complicate this link: the growth of core mass through gas inflows \citep[e.g.,][]{csengeri2014, alvarez2024, sandoval2025, beuther2025} and the subfragmentation of cores \citep[e.g.,][]{broganHunter2016, olguin2022, yoo2025, luo2026}. 

More recently, the ALMA-IMF Large Program \citep{motte2022} discovered that young massive protoclusters -- including our present target, W43-MM1 -- had top-heavy CMFs, meaning they had an excess of high-mass cores relative to low-mass cores \citep[][Cunningham et al. in prep.]{pouteau2022, nony2023, louvet2024}. These results are consistent with those of the ALMA-IMF pilot program \citep{motte2018b} and other large programs \citep{coletta2025, morii2026}, which call into question the universality of the CMF. Regarding the IMF, for a long time its shape has not been observed to vary much, so the community assumed it was universal \citep{bastian2010, offner2014}. However, some top-heavy IMFs have recently been observed in young massive clusters \citep[see][]{lu2013, maia2016, schneider2018, hosek2019} and some theoretical arguments favor the non-universality of the IMF \citep{kroupa2013, hopkins2018}. While the IMF of stars forming in the massive protoclusters observed with ALMA remains unknown, studying the effects of core subfragmentation and core mass growth on the CMF is crucial. Previous studies have relied on stellar multiplicity prescriptions \citep[e.g.,][]{swift2008, clarkWhitworth2021} that may not accurately depict core subfragmentation because stellar multiplicity is the result of the core evolution into stars followed by stellar dynamics.

It is challenging to account for the effect of core subfragmentation because it requires the use of stochastic properties measured in clouds to make deterministic predictions regarding how many star-forming fragments a core will host and the mass of each gas reservoir. Thus far, only a few studies have been dedicated to the cloud mass cascade, and they have barely constrained some of the properties of core fragmentation \citep[e.g.,][]{pokhrel2018, thomasson2022, beuther2025}. Concurrently, fragmentation studies conducted across two physical scales provided constraints that depend heavily on the selected scales and the completeness level of the observations, which restricts the investigation of physical effects on fragmentation \citep[e.g.,][]{bontemps2010, palau2013, traficante2023, morii2024}. The present study uses the multi-resolution database constructed for the W43-MM1 protocluster, which has a top-heavy CMF (see Sect.~\ref{s:data-cat}). With the \FAMILY hierarchy analysis tool presented in Sect.~\ref{s:family}, we characterize its  structure cascade, which range from above to below 2.4~kau, which is the cores' scale in W43-MM1 (see Sect.~\ref{s:results}). Then in Sect.~\ref{s:discussion}, we search for variations in the hierarchical mass cascade, compare the characteristics of various fragmentation studies, and predict the IMF that emerges from the top-heavy CMF of W43-MM1. Finally, Sect.~\ref{s:conc} summarizes the main findings of the paper.

\setcounter{figure}{0}
\begin{figure*}[htbp!]
    \centering
    \includegraphics[width=1.0\textwidth]{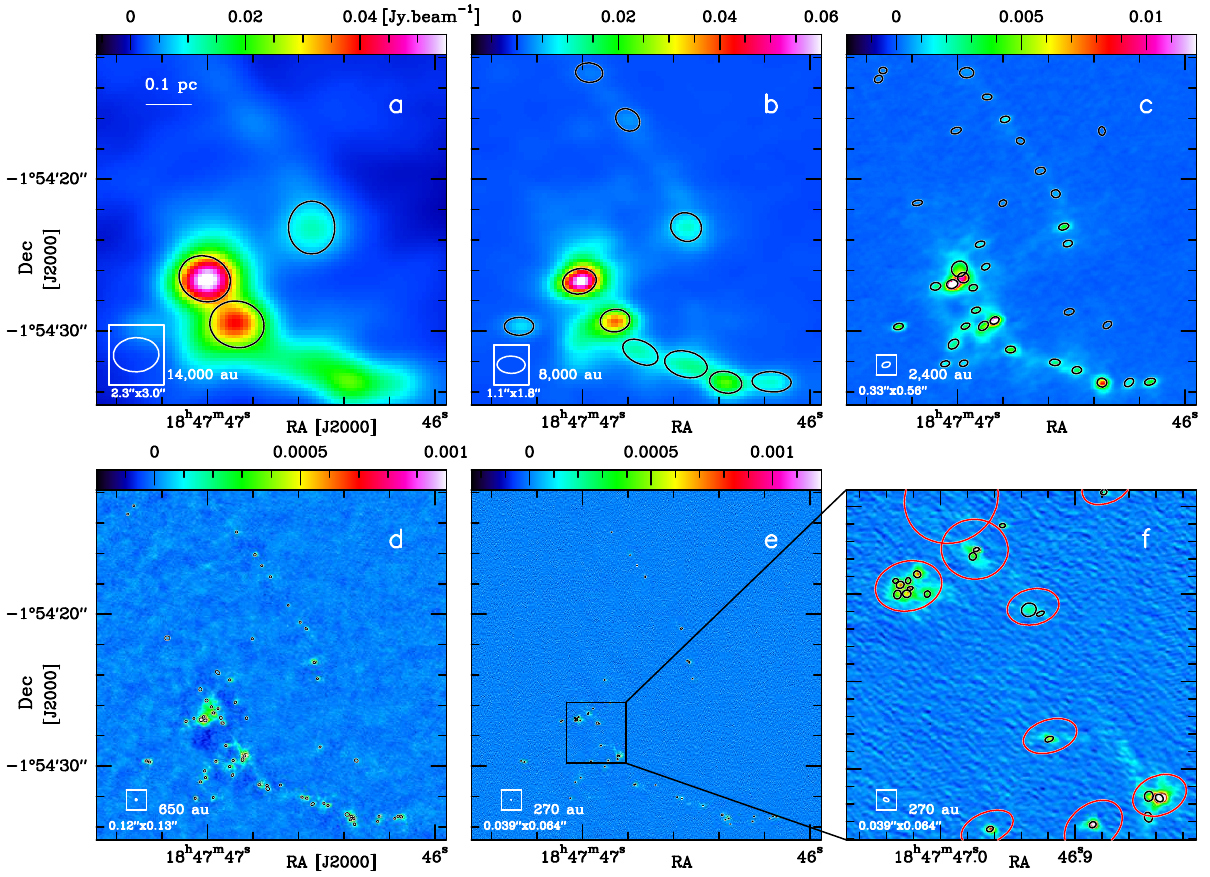}
    \vskip -0.1cm
    \caption{Central part of the W43-MM1 protocluster, as revealed by five 3~mm continuum ALMA images. \textit{Panels~a--e:} Images with resolutions ranging from $\sim$$2.6\arcsec$ to $\sim$$0.050\arcsec$, corresponding to 14~kau to 0.27~kau (see \cref{tab:sample}). Black ellipses outline the FWHM size of sources extracted from each image using \textsl{getsf}. A 0.1~pc scale bar is given in \textit{panel a}, and the beam size is shown in the bottom left corner of each image.
    \textit{Panel~f:} Zoom of the inner part of \textit{panel e}, with red and black ellipses outlining sources extracted at resolutions of 2.4~kau and 0.27~kau, respectively.}
    \label{fig:input-images-Main}
\end{figure*}

\begin{table*}[htbp!]
\centering
\begin{threeparttable}[c]
\caption{Database used for the hierarchical analysis of protoclusters with \textsl{FAMILY}.}
\label{tab:sample}
\begin{tabular}{llcllrc}
\hline \noalign {\smallskip}
Target/ &  Image    & Beam & \multicolumn{2}{c}{Resolution}   & Noise & Number of  \\
~~ Instrument    &   & [$\arcsec\times\arcsec$]  & [$\arcsec$] & [kau]   & [MJy\,sr$^{-1}$]  & extracted sources  \\
(1) & (2) & (3) & (4) & (5) & (6) & (7) \\
\hline \noalign {\smallskip}
W43-MM1/    & Cycle 6, natural weighting, $\sim$85~GHz  & $2.3\times3.0$    & 2.6   & 14   & $<$1.1 &3 \\
~~ALMA 12~m/ & Cycle 6, uniform weighting, $\sim$99~GHz & $1.1\times1.8$    & 1.45  & 8    & 2.5    &13 \\
~~ 3~mm (Band 3) & Cycle 5, robust 0, $\sim$101~GHz      & $0.33\times0.56$  & 0.43  & 2.4  & 8      &48 \\
            & Cycle 7, natural weighting, $\sim$85~GHz  & $0.12\times0.13$  & 0.12  & 0.65 & 35     &71\\
            & Cycle 7, uniform weighting, $\sim$99~GHz  &$0.039\times0.064$ & 0.050 & 0.27 & 450    &57 \\
\hline
Orion-like/ & Ideal MHD, X projection   &  72 & 72  & 40  & $0.25$ & 14\\
~~RAMSES                                && 36 & 36  & 20  & $0.32$ & 27\\
~~simulations/                          && 18 & 18  & 10  & $0.44$ & 49\\
~~synthetic map                         && 9  & 9   & 5   & $0.61$ & 63\\
\cline{2-7}
~~@ 3~mm 
& HD, Y projection                      & 72  & 72  & 40  & $0.40$ & 78\\
                                        && 36 & 36  & 20  & $0.47$ & 194\\
                                        && 18 & 18  & 10  & $0.62$ & 355\\
                                        && 9  & 9   & 5   & $0.78$ & 452\\
\hline \noalign {\smallskip}
\end{tabular}
(1)--(2) Basic parameters of the observational or synthetic images.
(3)--(4) Small and big axes of the ALMA half power beam width (Col.~3) and mean angular resolutions of the images (Col.~4).
(5) Spatial resolution adopting a distance of 5.5~kpc for W43 and assuming 550~pc for the synthetic clouds.
(6) Noise level estimated over low-intensity areas of the images, close to hierarchical structures. 
(7) Number of compact sources extracted with the \getsf algorithm.
\end{threeparttable}
\end{table*}

\section{Database: images and source catalogs}  
\label{s:data-cat}

The present study makes use of five images of the W43-MM1 protocluster, which trace its cloud structure from $\sim$14~kau to $\sim$0.27~kau (see Sects.~\ref{s:obs}--\ref{s:catalog}). For testing purposes, four images of two synthetic protoclusters from numerical simulations are jointly analyzed (see Sect.~\ref{s:simu}). Figures~\ref{fig:input-images-Main} and \ref{appendixfig:input-images-SW} display the five W43-MM1 images that we used in the present analysis.

\subsection{Data reduction of the ALMA images targeting W43-MM1}
\label{s:obs}

We used three different 3~mm datasets obtained with ALMA in different array configurations, and each of these datasets covering the central 1.6~pc\,$\times$\,1.6~pc part of the W43-MM1 protocluster. The first dataset comes from the ALMA-IMF\footnote{
    See \url{https://alma-imf.cnrs.fr}.  
    Data available on \url{https://zenodo.org/record/5598066}.}
Large Program, which produced a large mosaic of the W43-MM1 protocluster in Cycle~5, with a spectral coverage of $91.7-105.5$~GHz \citep[\#2017.1.01355.L, PIs: Motte, Ginsburg, Louvet, Sanhueza, see][Paper~I]{motte2022}. The other two datasets correspond to a single pointing of $\sim$$60\arcsec$, centered on the brightest ALMA-IMF cores with coverage at 3~mm of $84.1-99.9$~GHz: one from ALMA Cycle~6 and the other from ALMA Cycle~7 (\#2018.1.01787.S and \#2019.1.00502.S, PI: Louvet). We used the 3~mm continuum image of ALMA-IMF, which was reduced using CASA\footnote{
    \url{https://casa.nrao.edu}} 
as described in companion paper, Paper~II \citep{ginsburg2022}. With a robust parameter of 0, the mean resulting angular resolution is $\sim$0.43$\arcsec$, corresponding to $\sim$2.4~kau, adopting a distance of 5.5$\pm$0.4~kpc \citep{zhang2014}. 

With the same attention to sensitivity, angular resolution, and line contamination, the Cycle~6 and Cycle~7 data were processed with IMAGER\footnote{
    IMAGER is an interferometric imaging package in the GILDAS software (\url{http://www.iram.fr/IRAMFR/GILDAS}), tailored for usage simplicity and efficiency for multi-spectral datasets. See \url{https://imager.oasu.u-bordeaux.fr}.}.
We used the two continuum bands (bandwidth of 1.875~GHz centered at $\sim$85~GHz and $\sim$99~GHz) with a filter to remove the regions of the spectrum contaminated by spectral lines above the $1\sigma$ noise level (typically 50~mK and 4-5~K for Cycle~6 and Cycle~7 data, respectively). We applied phase and amplitude self-calibration and the cleaning was performed using Clark deconvolution. With natural and uniform weightings (corresponding to robust 2 and $-2$ parameters) on the $\sim$85~GHz and $\sim$99~GHz images, respectively, Cycle~6 data provided images with a mean spatial resolution of $\sim$2.6$\arcsec$ and $\sim$1.45$\arcsec$ (i.e., $\sim$14~kau and $\sim$8~kau), respectively. With the same weightings and frequencies, Cycle~7 images have a $\sim$0.12$\arcsec$ and $\sim$0.050$\arcsec$ (i.e., $\sim$650~au and $\sim$270~au) resolution, respectively. The basic characteristics of the five observational images, including their spatial resolution and noise levels, are listed in \cref{tab:sample}. 

For this project, which probes very dense structures at the highest angular resolutions, the most suitable wavelength in young protoclusters like W43-MM1 is 3~mm. Indeed, the 3~mm emission is largely dominated by dust thermal  emission, rather than free-free, and more likely remains optically thin than at 1~mm.  Consistent with this, companion paper, Paper~XX \citep{yoo2025} showed that the 1~mm emission of the W51-E and W51-IRS2 protoclusters, which were studied at $\sim$200~au scales, is partly optically thick. We refrain from creating images at intermediate scales by convolving some of the observed images because our goal is to keep them as independent of each other as possible. The maximum recoverable scale of the current dataset, which is about ten times the natural-weighting beam, is not an issue since our objective is to identify compact sources with full width at half maximum (FWHM) sizes of one to three times the beam (see Sect.~\ref{s:catalog}).

\subsection{Compact sources in W43-MM1} 
\label{s:catalog}

Since our goal is to extract sources with different characteristic sizes in each image, we chose to use a software package that identifies and characterizes emission peaks: \getsf \citep{men2021getsf}. The size of the extracted sources is limited by their structured background and neighboring sources. Therefore, as discussed in \cite{louvet2021simu}, the size distribution and thus the flux distribution of sources is directly related to the angular resolution of the image. Using this property, \getsf has already been used to characterize the hierarchical structure of a cloud associated with a protocluster \citep{thomasson2022}. We used \getsf (v240325) for each of the five images of W43-MM1 with angular resolutions of 2.6\arcsec, 1.45\arcsec, 0.43\arcsec, 0.12\arcsec, and 0.050\arcsec.

In detail, the \getsf software package employs a spatial decomposition of the observed images to better isolate various spatial scales and separate the structural components of relatively round sources and elongated filaments from each other and from the background. This method has a single free parameter, the maximum size of the sources to be extracted. In order to focus on compact sources, we kept this parameter small, that is three times the beam of each image. The detection provides a first-order estimate of source footprints, sizes, and fluxes. As a second step, robust measurements of the sizes and fluxes of sources are made on background-subtracted images. The resulting catalog contains, for each source, its coordinates, the parameters of its Gaussian FWHM elliptical size, its peak, and integrated fluxes. \getsf extracted between 3 to 71 sources in the W43-MM1 images at 3~mm (see \cref{tab:sample}). The number of sources is lower in the lowest angular resolution images, that is those from Cycle~6, because all of our input images cover the same spatial area. The number of sources is also lower in images that were cleaned with uniform weighting because the angular resolution is improved at the expense of sensitivity. That is particularly obvious for the best-angular Cycle~7 image, which has a resolution of $0.050\arcsec$ (see \cref{tab:sample}). We will address the issue arising from the unbalanced completeness of our catalogs in Sect.~\ref{s:results}. Although the W43-MM1 individual images have been observed and reduced differently, interferometric filtering is not expected to significantly affect the flux estimates of compact sources extracted in W43-MM1 by \getsf \citep[see, e.g., Appendix~A of][]{pouteau2022}. \cref{tab:MM1sources} lists the characteristics of each source extracted by \textsl{getsf}.

The images targeting the central part of the W43-MM1 protocluster have spatial resolutions of 14~kau, 8~kau, 2.4~kau, 0.65~au, and 0.27~au. In the following, we will call `clumps' sources extracted in the two lowest angular resolution images of Cycle~6 because their typical deconvolved sizes range from 5~kau to 30~kau is larger than typical core sizes in massive protoclusters \citep[see][]{motte2018a}. The sources of the ALMA-IMF 3~mm image have the size of cores, as proposed by previous ALMA-IMF papers \citep[e.g., Papers~I, III and XV,][]{motte2022, pouteau2022, louvet2024}. We refer as `fragments' the sources extracted in the highest spatial resolution images of Cycle~7 because they have typical deconvolved size range from 1000~au to 150~au. They are quoted as pre/protostellar objects (PPOs) by \cite{yoo2025} when identified in images at $100-300$~au resolutions. We will use the term `seeds' to refer to fragments identified at the 200~au scale, at which point rotation becomes dominant. In the present paper, we assume that protostellar fragments remain dynamically coupled to their parental cloud hierarchy. This occurs in numerical simulations when a cloud is forming and stellar dynamics are not yet significant \citep{bernard2025}. This situation is highly probable in the W43-MM1 protocluster because it is young, and its cloud is still collapsing globally \citep{motte2003, nguyen2013}.

The prestellar versus protostellar nature of the sources extracted in these five 3~mm images is not easily defined. In particular, the Cycle~7 dataset is not sensitive enough (see Sect.~\ref{s:obs}) to conduct a thorough survey of protostellar outflows or trace many hot core lines. We can only rely on the ALMA-IMF outflow surveys \citep[including Papers~V, X, and XVII,][Nony et al. in prep.]{nony2020, nony2023, armante2024, valeille2025} and hot core surveys \citep[Papers~IV and XI,][]{brouillet2022, bonfand2024}, which were performed at the core scale. Since the determination of the prestellar or protostellar nature of the sources is incomplete in the five W43-MM1 catalogs, we will not attempt below to use this characterization to separate the structures according to their nature.

\subsection{Synthetic compact source catalogs}
\label{s:simu}

For test and initial comparison purposes, we used source catalogs derived from synthetic images of numerical simulations designed to represent the Orion protocluster \citep[$10^5~\msun$ within $\sim$(20~pc)$^3$,][]{louvet2021simu}. The formation of this protocluster was simulated by solving the magneto-hydrodynamical (MHD) or hydrodynamical (HD) equations with the RAMSES code \citep[see][]{ntormousi2019}. An isothermal equation of state was used, as well as no recipes for stellar feedback. \cite{louvet2021simu} simulated the $250\,\mu$m observation at  $9\arcsec-72\arcsec$ resolutions of a cloud located at 1.75~kpc. We rescaled these observations to those of an imaginary cloud located at 550~pc. This allows us to study smaller physical scales than in \cite{louvet2021simu} while remaining at scales larger than 5~kau, where the ideal MHD and an isothermal equation of state are valid. We caution that these simulations should not be used to compare mass cascades below the core scale. We also changed the synthetic observation frequency to remain in the optical thinness regime assumed by \cite{louvet2021simu}. To simulate the 3~mm emission of these synthetic clouds, we divided their $250\,\mu$m fluxes by a factor of $1.2\times 10^3$, corresponding to the flux ratio expected for a 20~K modified blackbody with a dust emissivity law of $\kappa(\nu)\propto \nu^{1.5}$.

Using \textsl{getsf}, \cite{louvet2021simu} extracted between 14 to 452 sources from these two sets of synthetic images (see \cref{tab:sample}). The HD simulations of \cite{ntormousi2019} provide a larger number of compact sources, $\sim$7 times more than MHD simulations \citep{louvet2021simu}, but they are less suited to represent the complex physics of the W43-MM1 massive protocluster. The synthetic source catalogs are employed here to characterize the fragmentation cascade above the scale of cores and to verify the reliability of W43-MM1 measurements conducted with limited statistics. Numerical simulations capturing the mass concentration in W43-MM1, solving non-ideal MHD equations, using radiative transfer, and including stellar feedback will be necessary to probe smaller spatial scales.

\begin{figure}[htbp!]
    \centering
    \includegraphics[width=0.35\textwidth]{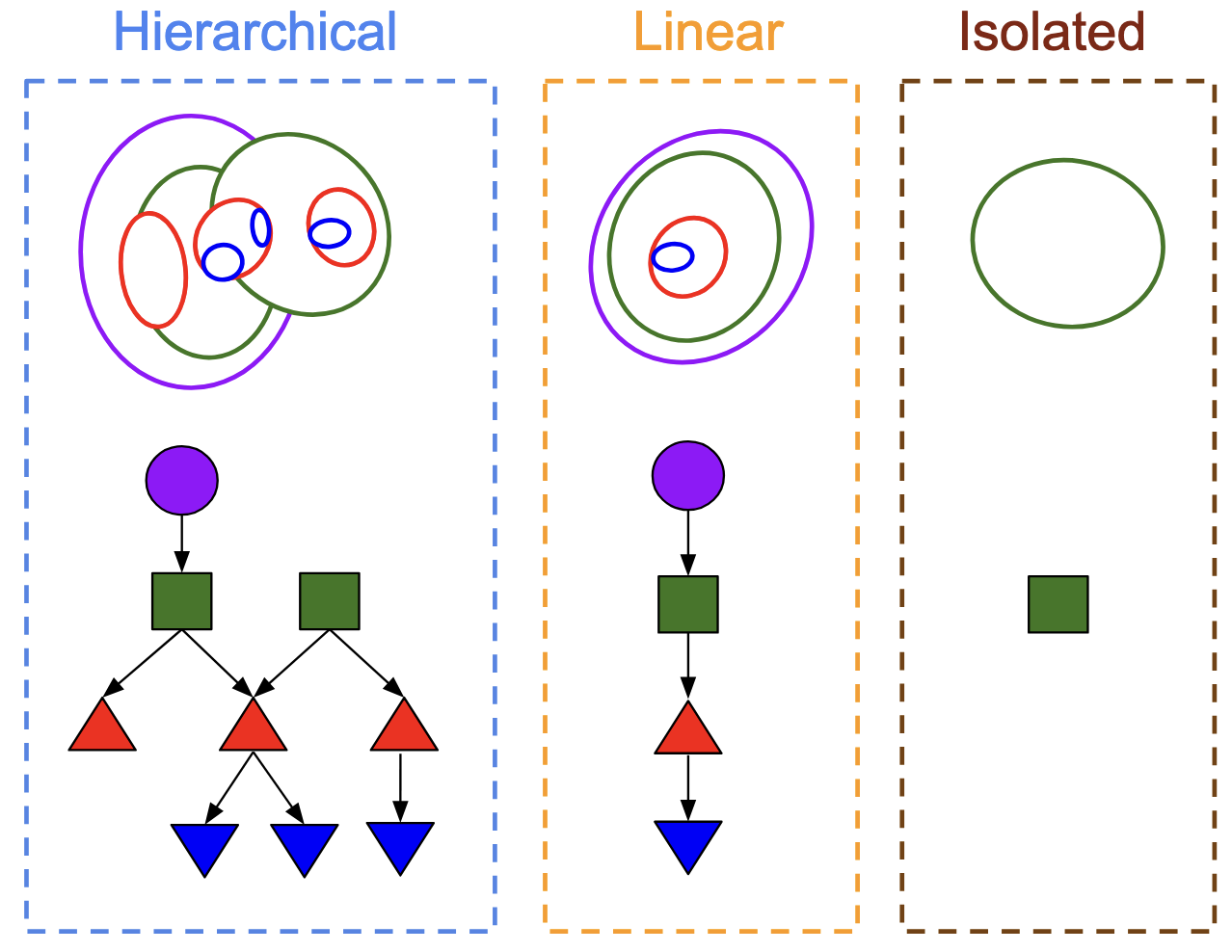}
    \caption{The three types of structures as defined by \textsl{FAMILY}. 
    \textit{Top row:} Stack of sources detected at four increasing resolutions (purple, green, red, and blue figures). \textit{Bottom rows:} Multiscale networks of sources qualified either as hierarchical (\textit{left box}, several children at least at one level), or linear (\textit{central box}, a single child source with at most one parent source at each level), or isolated (\textit{right box}, no child or parent sources). Most W43-MM1 sources are part of hierarchical structures.}
    \label{fig:struct-types}
\end{figure}

\begin{figure}[htbp!]
    \hskip -0.2cm \includegraphics[width=0.5\textwidth]{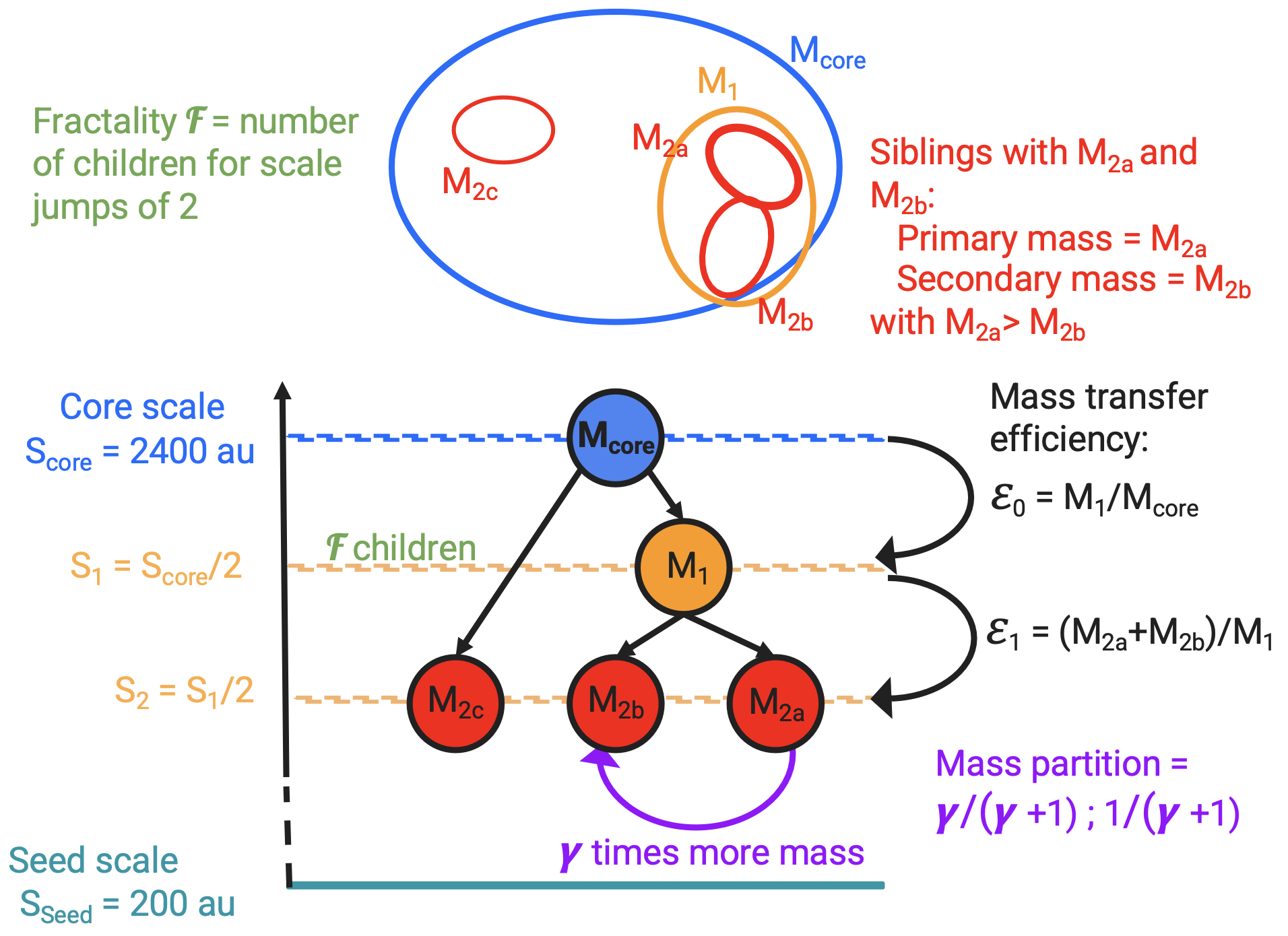}
    \vskip -0.1cm
    \caption{Scheme of the hierarchical fragmentation cascade from the core to the seed scales that highlights its parameters. Stack sources (\textit{upper panel}) and representation of the hierarchical structure (\textit{lower panel}). For each decrease of a constant factor (here two) in physical scales, the fractality index is represented by $\mathcal{F}$ and the mass transfer efficiency by $\epsilon$. For each sibling, the mass ratio between the primary and secondary children, $\gamma$, gives their mass partition.}
    \label{fig:param-definition}
\end{figure}

\section{The multiscale network analysis tool \textsl{FAMILY}}
\label{s:family}

The exact shape of the IMF emerging from the CMF depends on the fragmentation process within cores, which is difficult to constrain observationally. According to \cite{joncour2018} and \cite{thomasson2026}, the recipes for the subfragmentation of cores can be extrapolated from the structure cascade measured at scales larger than the core scale. Measuring the parameters of this cascade in detail is then the first necessary step in predicting the emergent IMF \citep{thomasson2026}. 

\cite{thomasson2022} developed a graph-theory-based methodology to represent the multiscale structure of the cloud as a network of nested compact sources. The associated analysis tool is hereafter called \textsl{FAMILY}\footnote{
    \url{https://github.com/ThomassonB/FAMILY}},
which stands for Framework for Analyzing MultIscaLe sYstems. \FAMILY establishes a parental link between sources that are identified at increasing spatial resolutions and characterized by Gaussian parameters, in particular their peak location and their major and minor FWHM sizes. At least three catalogs of sources detected at different spatial resolutions are required. Each source is assigned a so-called \FAMILY node number (see \cref{tab:MM1sources}). The only free parameter of this method is the minimum common area between related sources, which are assimilated as ellipses with $\frac{2\times \rm FWHM}{\sqrt{2 \ln 2}}$ outer diameters. We hereafter set this overlap coefficient to $75\%$ of the child area as proposed by \cite{thomasson2022}. Figure~\ref{fig:struct-types} classifies structures with several children at least at one level as `hierarchical', calls those with a unique child at least at two fragmentation levels as `linear', and quotes those without any parental or child link as `isolated'.

\FAMILY provides all the necessary characteristics of a cloud's mass cascade, represented by a fractal network over a given range of scales: the fractality index, mass transfer efficiency, and mass partition between siblings \citep{thomasson2022, thomasson2024, thomasson2026}. From the entire set of sources extracted at different physical scales, \FAMILY computes the fractality index, $\mathcal{F}$, as the value at which an ideal fractal cascade matches the total number of sources \citep{thomasson2022}. The fractality quantifies the average number of fragments hosted by a cloud structure when the scale is reduced by a constant factor, which is set to two in the present study. Figure~\ref{fig:param-definition} illustrates a multiscale network and describes the parameters associated with the equivalent scale-free hierarchical fragmentation from the core to the seed scales. The structure of \cref{fig:param-definition} has five nodes, which are distributed across three evenly spaced levels by a factor of two. According to Eq.~9 from \cite{thomasson2022}, an ideal fractal model yields the equation $5 = 1 + \mathcal{F} + \mathcal{F}^2$, whose positive root solution is the average fractality index of $\mathcal{F} = 1.56$. Based on Eq.~4 from \cite{thomasson2022}, the fractality index $\mathcal{F}$ enables the computation of the multiplicity level of a parental structure, which is defined as the number of fragments it contains. Notably, a core detected at a scale of $s_{\rm core}=2400$~au will fragment into $n_{\rm seed}$ seeds at the seed scale of $s_{\rm seed}=200$~au. $n_{\rm seed}$ is computed as follows:
\begin{equation}
     n_{\rm seed} = \left(\frac{s_{\rm core}}{s_{\rm seed}}\right)
     ^{\frac{\ln\mathcal{F}}{\ln 2}} = \left(\frac{\rm 2400~au}{\rm 200~au}\right)
     ^{\frac{\ln\mathcal{F}}{\ln 2}}.
    \label{eq:fractality}
\end{equation}

\begin{table*}[t!]
\centering
\resizebox{\textwidth}{!}{
\begin{threeparttable}[c]
\caption{Characteristics of the mass cascades in W43-MM1 and in synthetic protoclusters, compared to the initial study of NGC~2264.}
\label{tab:fragm-casc_measured}
\begin{tabular}{lccccccccc}
\hline \noalign{\smallskip}
Protocluster & Scale range & \multicolumn{3}{c}{Number of structures} & $\mathcal{F}_{\rm 2D}$ & $\mathcal{F}_{\rm 3D}$ & $\overline{\epsilon_{\rm jump2}}$ & $\overline{\gamma}$ & Mass\\
       & [kau] & hierarchical & linear & isolated &        &  & [\%] &  & partition\\
(1) & (2)    & (3) & (4) & (5)                          & (6) &  (7)               & (8)  & (9) & (10) \\
\hline \noalign{\smallskip}
W43-MM1 & $0.27-14$ & 9 (72\%) & 18 (24\%) & 7 (4\%) & $1.17\pm0.02$ & $1.19\pm0.10$ &  $\gtrsim$60 &  $\sim$1.95 & [66\%; 34\%]\\
Orion-like MHD & $5-40$  & 5 (53\%) & 24 (39\%) & 12 (8\%) & $1.39\pm0.09$ & $1.47\pm0.15$ &  $60\pm25$ &  $\sim$1.56 & [61\%; 39\%]\\
Orion-like HD & $5-40$ & 92 (64\%) & 120 (28\%) & 87 (8\%) & $1.35\pm 0.06$ & $1.43\pm0.15$ & $65\pm25$ & $\sim$1.42 & [59\%; 41\%] \\
\hline
NGC~2264 & $1.4-26$  & 23 (31\%) & 135 (47\%) & 176 (22\%) & $1.45\pm0.12$ & $1.70\pm0.24$  &  $-$ &  $-$ & $-$ \\
\hline \noalign{\smallskip}
\end{tabular}
\begin{tablenotes}
\item{(1)--(2)} Protoclusters (Col.~1) studied over a given range of physical scales (Col.~2). The result of the initial study of NGC~2264 \citep{thomasson2022, thomasson2024} is given for comparison.
\item{(3)--(5)} Number of structures identified by \FAMILY (see definitions in Sect.~\ref{s:family} and \cref{fig:struct-types}). The percentage of source numbers contained within each structure type is provided in parentheses. 
\item{(6)--(7)} Median 2D fractality index of hierarchical structures (Col.~6). 3D fractality index (Col.~7) estimated by correcting $\mathcal{F}_{\rm 2D}$ for the projection effect, which depends on the jumps in physical scales \citep[see][and Figs.~\ref{fig:param-definition} and \ref{appendixfig:3Dfractality}]{thomasson2022, thomasson2024}.
\item{(8)} Median mass transfer efficiency for a jump in physical scales of two, as defined in Sect.~\ref{s:family} (see also \cref{fig:param-definition}). For reasons given in Sect.~\ref{s:MassTransfer}, $\overline{\epsilon_{\rm jump2}}$ is equal to the flux transfer efficiency in the present synthetic protoclusters and represents a lower limit for W43-MM1.
\item{(9)} Median value of the mass ratios of the dominant (primary) child to the secondary child of the same parental structure, as defined in Sect.~\ref{s:family} (see also \cref{fig:param-definition}).
\item{(10)} Median value of the mass partition between the primary and secondary children of the same parental structure, computed from Col.~9. As explained in Sect.~\ref{s:MassTransfer}, the mass partition  of binary fragments is equal to their flux partition in present synthetic protoclusters, and it corresponds well to the flux partition in W43-MM1.
\end{tablenotes}
\end{threeparttable}
}
\end{table*}

In \cref{fig:param-definition}, the mass transfer efficiency is the ratio of the sum of the masses of child sources, $M_{\rm child}$, to the mass of their direct parental source, $M_{\rm parent}$: $\epsilon=\sum M_{\rm child}/M_{\rm parent}$. Assuming this efficiency does not vary with spatial scale, source density, or across the cloud, one can compute its median value by considering all the parent-children pairs in the linear and hierarchical structures \citep{thomasson2024}. For a jump of two in physical scales, this median mass transfer efficiency, $\overline{\epsilon_{\rm jump2}}$,  allows one to calculate the formation efficiency of seeds at the smallest observed scale within a core:
\begin{equation}
    \epsilon^{\rm core}_{\rm seed} = \left(\overline{\epsilon_{\rm jump2}}\right)^\eta, {\rm with~} \eta=
    {\frac{\ln \frac{s_{\rm core}}{s_{\rm seed}}}{\ln 2}}.
    \label{eq:eff}
\end{equation}
Following the definition of stellar systems, \cite{thomasson2026} introduced the mass partition between siblings of fragments from the same parental cloud structure (see \cref{fig:param-definition}). When there are two children sources in a given parental structure, the ratio of the most massive child mass, $M_{\rm primary}$, to that of the least massive, $M_{\rm secondary}$, is $\gamma=\frac{M_{\rm primary}}{M_{\rm secondary}}$. Using the median value of this ratio computed over all pairs of sources, $\overline{\gamma}$, we define the mass partition, $\rm MP(Primary;Secondary)$, as follows:
\begin{equation}
    \rm MP(Primary;Secondary) = \left[\frac{\overline{\gamma}}{\overline{\gamma}+1}; \frac{1}{\overline{\gamma}+1}\right].
    \label{eq:partition}
\end{equation}

As shown by \cite{thomasson2024}, the fractality index and mass transfer efficiency characterize the fragmentation regime of the cloud from turbulence-dominated to gravo-thermally dominated. They developed a stochastic gravo-turbulent model that simulates the fragmentation cascade in a star-forming cloud. For this model to be predictive, input parameters derived from the fractality index, mass transfer efficiency, and mass partition must be correctly constrained. This will be done using \textsl{FAMILY}.

\begin{figure}[htbp!]
    \centering
    \vskip -0.4cm
    \includegraphics[width=0.4\textwidth]{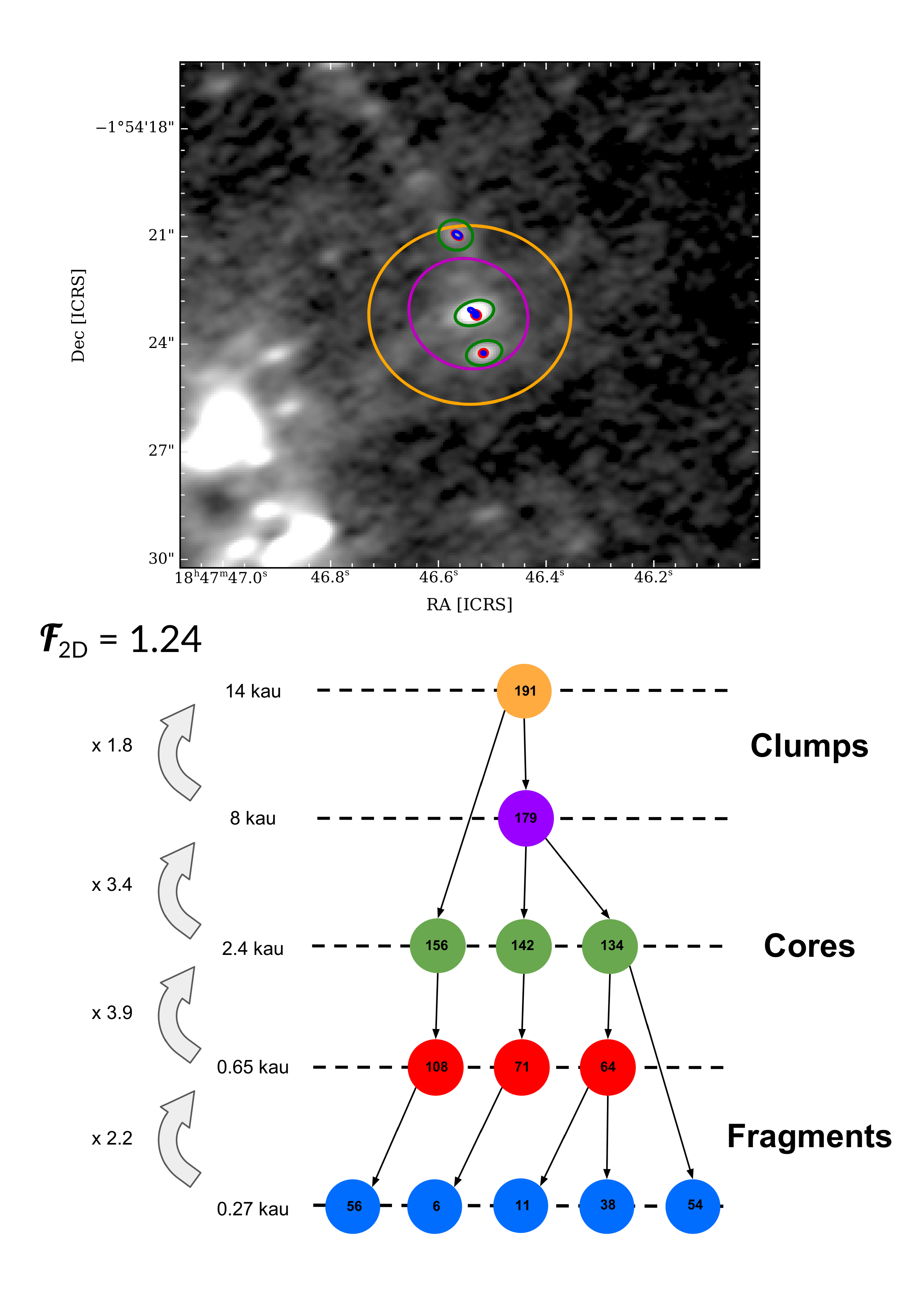}
    \vskip -0.5cm
    \caption{Fourth richest hierarchical structure of the W43-MM1 protocluster with a fractality index of $\mathcal{F}_{\rm 2D}\simeq1.24$. \textit{Upper panel:} Stack sources identified at clump resolutions of 14~kau (yellow), 8~kau (purple), at the core scale of 2.4~kau (green), and at fragment resolutions of 650~au (red) and 270~au (blue). The underlying image is the 3~mm continuum emission at $0.43\arcsec$ resolution. Ellipses here represent the outer diameter of the sources, which is $\sim$1.7 times the FWHM of the ellipses shown in Figs.~\ref{fig:input-images-Main} and \ref{appendixfig:input-images-SW} (see Sect.~\ref{s:family}). 
    \textit{Lower panel:} Representation of the associated hierarchical structure, showing the relationship established between sources of each scale, here identified by their node number (see also \cref{tab:MM1sources}).
    }
    \label{fig:MM1-ExampleHierarchical}
\end{figure}

\section{Characterization of the hierarchical mass cascade}
\label{s:results}

Here, we use \FAMILY to analyze the hierarchical mass cascade of the W43-MM1 protocluster and the two synthetic protoclusters. We identify their multiscale network of sources in Sect.~\ref{s:network} and characterize their cascade parameters in Sects.~\ref{s:fractality}--\ref{s:MassTransfer}. \cref{tab:fragm-casc_measured} lists all these parameters for W43-MM1 and the synthetic protoclusters of Sect.~\ref{s:simu}, along with those estimated by \cite{thomasson2022, thomasson2024} for the NGC~2264 protocluster.

\subsection{Networks of nested sources}
\label{s:network}
We applied the \FAMILY analysis tool of \cite{thomasson2022} to create a parental link between the compact sources of each W43-MM1 catalog and of each catalog of the two synthetic protoclusters, respectively. \cref{tab:MM1sources} is complemented by the source list corresponding to the ascending \FAMILY tree of each W43-MM1 source detected in Sect.~\ref{s:catalog}.

In W43-MM1, we defined 34 structures, including nine hierarchical, 18 linear, and seven isolated (see \cref{tab:fragm-casc_measured}). Figures~\ref{fig:MM1-ExampleHierarchical} and \ref{appendixfig:Hierarchical} display the nine hierarchical structures of W43-MM1 and provide a representation of their source network. The hierarchical structures of W43-MM1 contain 3 to 71 sources. The main hierarchical structure of W43 contains $\sim$40\% of the sources discovered in this massive protocluster and presents characteristics that dominate those of the entire protocluster. Figure~\ref{appendixfig:spatial-distrib} displays the spatial distribution of all structures identified in W43-MM1 compared to those of the MHD synthetic protocluster. The W43-MM1 protocluster consists of a cluster of rich, hierarchical structures associated with the main clump, sometimes called a ridge or hub \citep{nguyen2013}. In numerical simulations, hierarchical structures are located along the densest filaments.

\subsection{Fractality index of the hierarchical cascade} 
\label{s:fractality}

The fractality index, whose definition is given in Sect.~\ref{s:family}, theoretically differs in regimes dominated by incompressible turbulence, or by other phenomena such as gravity and stellar feedback \citep{thomasson2024}. Since gravity is expected to be the most significant phenomenon at the scales observed in W43-MM1, we first estimated a single fractality index. Due to the limited number of sources observed in W43-MM1, we refrain from evaluating how fractality varies with respect to the environment, the mass, or nature of the sources. In the following, we therefore assume that the hierarchical cascade is homogeneous and fully self-similar; we will start revising this assumption in Sect.~\ref{s:cascade-var}. We used the structures identified in Sect.~\ref{s:network} to compute the fractality index in both the W43-MM1 and the synthetic protoclusters. This index is calculated as the median value measured for all hierarchical structures. Since isolated plus linear structures account for only $\sim$28\% of the sources in W43-MM1 (see \cref{tab:fragm-casc_measured}), the fractality index accurately describes the overall distribution of sources within the cloud structures of this observed protocluster. The situation is less clear for the synthetic MHD and observed NGC~2264 protoclusters, whose isolated plus linear structures contain $\sim$47\% and $\sim$69\% of the sources, respectively (see \cref{tab:fragm-casc_measured}).

\begin{figure}[htbp!]
    \centering
    \includegraphics[width=0.25\textwidth]{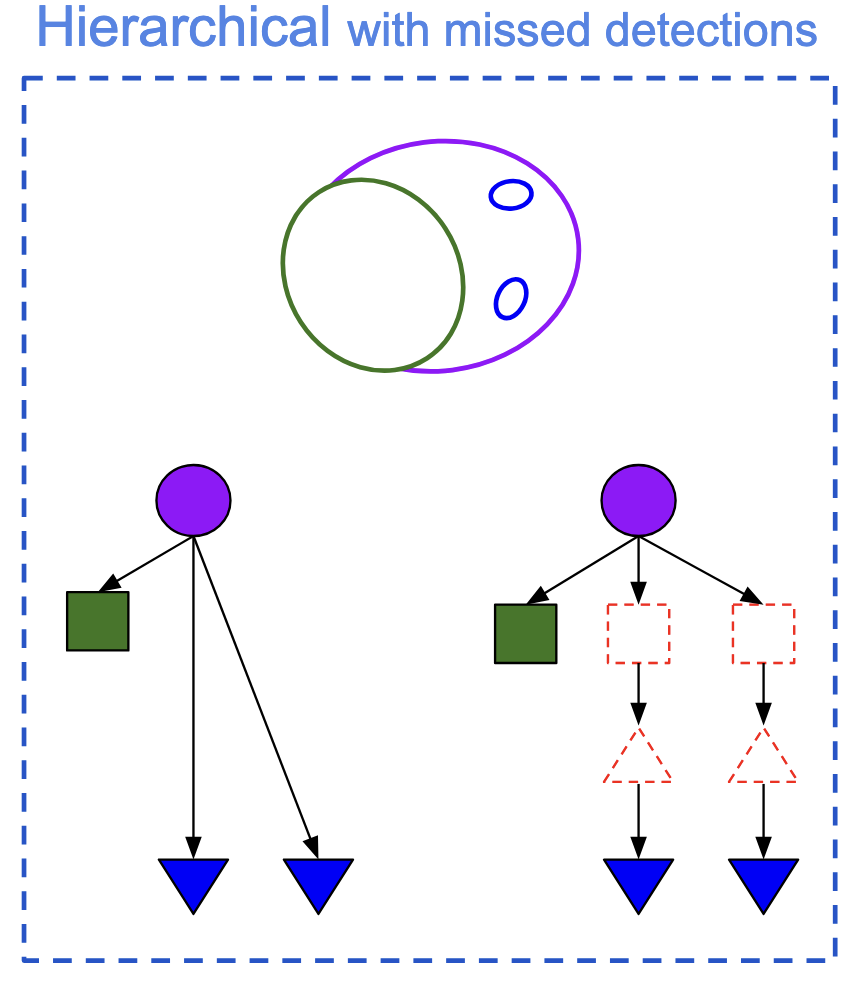}
    \caption{2D fractality index of hierarchical networks, estimated as the average of the fractalities measured with and without considering undetected sources at certain scales (\textit{right} and \textit{left structures}, see \citealt{thomasson2022}). The lower limit is obtained without any addition. Missed detections added to the network (open dotted symbols in the \textit{right structure}) result in a higher fractality value, which is used as an upper limit.}    
    \label{fig:fractality-missing}
\end{figure}

For the W43-MM1 protocluster as a whole, we measured the median 2D fractality index of $\mathcal{F}_{\rm 2D}$ to be between 1.167 to 1.172. The minimum value corresponds to that obtained from all existing sources in the network defined in Sect.~\ref{s:network}. The maximum value proposed by \cite{thomasson2022} is obtained by introducing virtual sources corresponding to those missing between two scales into the structure cascade (see \cref{fig:fractality-missing}). This serves to correct for the detection limits of a cloud hierarchy suggested by existing sources at a lower level. Since it assumes the existence of fragments at scales where a hierarchy may not yet exist, it is indeed an upper limit. Like \cite{thomasson2022}, we decided to take as the 2D fractality of the W43-MM1 protocluster the average value between these upper and lower limits: $\mathcal{F}_{\rm 2D}\simeq 1.170$. Given the small number of estimated missing sources in the \textsl{FAMILY}-detected stack levels, the $\sim$0.006 difference between the upper and lower limits is too small to represent twice the uncertainty of this index. However, the $\sigma \simeq 0.16$ dispersion of the fractality index across the hierarchical structures reflects physical variations and thus overestimates our measurement uncertainty. We therefore set this uncertainty to $0.02$ and estimate the fractality index of W43-MM1 to be $\mathcal{F}_{\rm 2D}\simeq 1.17\pm 0.02$. In the hierarchy of synthetic images obtained from MHD simulations, we measured a 2D fractality index of $\mathcal{F}_{\rm 2D} \simeq 1.39\pm 0.09$.
The 3D fractality indices, $\mathcal{F}_{\rm 3D}$, can be derived from $\mathcal{F}_{\rm 2D}$ by accounting for the projection effect that would dissimulate a source along the line of sight of another. We followed the methodology proposed by \cite{thomasson2024} to correct for this effect. Due to the jumps in physical scales that range from 1.8 to 3.9 for the W43-MM1 protocluster (see \cref{fig:MM1-ExampleHierarchical}), and is two for the synthetic observations of numerical simulations (see \cref{tab:sample}), we estimated that these fractality indices are underestimated by approximately $\sim$6\% (see \cref{appendixfig:3Dfractality}). The resulting 3D fractality indices are $\mathcal{F}_{\rm 3D}=1.19$ for the W43-MM1 protocluster and $\mathcal{F}_{\rm 3D}=1.47$ for the synthetic MHD protocluster.

We conducted tests to check that the W43-MM1 fractality measurement is robust against sensitivity biases. In particular, source catalogs derived from ALMA data tend to be noisier at higher angular resolutions (see Sect.~\ref{s:obs} and \cref{tab:sample}). Consequently, sources are particularly missing at the 270~au scale. Therefore, we examined the impact of fewer sources being detected at different scales by using the rich catalogs of sources extracted from synthetic images of HD simulations. We halved one of the four catalogs of \cref{tab:sample} four times, removing the faintest sources from each modified catalog. Then, we computed the fractality index associated with the resulting set of four catalogs and found that it varied by less than 17\%. Besides, we created W43-MM1 catalogs with equivalently low sensitivity at all scales by cutting the catalogs of \cref{tab:sample}, thereby removing the faintest sources at scales greater than 270~au. We found that the fractality index remains within $3\%$ of what was obtained with the original catalogs. 

Therefore, the average fractality index of W43-MM1 is robustly measured to be $\mathcal{F}_{\rm 3D}\simeq1.19\pm 0.10$. From \cref{eq:fractality} and $\mathcal{F}_{\rm 3D}=1.19$, it follows that, on average, a W43-MM1 core detected at a scale of $s_{\rm core}=2400$~au will host $n=2$ fragments at a scale $s_{\rm 2\,fragments}$, the value of which is computed as follows:
\begin{equation}
    s_{\rm 2\,fragments} =s_{\rm core}\times n^{\frac{\ln\mathcal{F}_{\rm 3D}}{\ln 2}}= 2400~\rm au \times 2^{\frac{\ln 1.19}{\ln 2}}
\simeq 150~{\rm au}.
    \label{eq:scale2}
\end{equation}

\subsection{Mass transfer through the structure cascade}
\label{s:MassTransfer}

Beyond the fractality index measured in Sect.~\ref{s:fractality}, characterizing the mass cascade requires estimating two parameters defined in Sect.~\ref{s:family} (see also \cref{fig:param-definition}): the mass transfer efficiency, $\overline{\epsilon_{\rm jump2}}$, and the mass partition, MP(Primary; Secondary). As demonstrated by \cite{thomasson2026}, these parameters are among the critical ones for the evolution of the CMF into an IMF.

The source catalogs of \cref{tab:MM1sources} provide source fluxes at varying frequencies, rather than masses. For a theoretical cloud with constant dust temperature and emissivity, as well as optically thin emission at the same frequency at all probed scales, the source fluxes are proportional to the source masses. This is the case for the synthetic catalogs of Sect.~\ref{s:simu} because the radiative transfer applied by \cite{louvet2021simu} to the large-scale cloud structures traced in the numerical simulations of \cite{ntormousi2019} makes these simple assumptions. Therefore, the distributions of flux transfer efficiency and flux partition that we measure for these simulations perfectly correspond to the distributions of mass transfer efficiency and mass partition. 
The situation is much more complex for real observations, such as those of Sect.~\ref{s:obs}, or for synthetic simulations tracing smaller physical scales, for several reasons. First, as shown by \cite{yoo2025}, sources detected at 100-300~au scales in massive protoclusters exhibit partially optically thick emission even at 3~mm. Second, dust emissivity is expected to vary from one frequency to another and from one source to another, particularly as a function of source density and temperature. Lastly for W43-MM1, even if source temperatures could be extrapolated from the relations proposed by companions papers, Papers~XII and XVI \citep{dellova2024, motte2025}, they would be highly uncertain for fragments whose prestellar versus protostellar nature has yet to be determined. Therefore, we consider the distributions of flux transfer efficiency and flux partition in the structure cascade of the W43-MM1 protocluster with utmost caution.

Figure~\ref{appendixfig:param-measured} shows the distributions of mass transfer efficiency directly computed from the flux transfer efficiency measured in the two synthetic protoclusters. Some values of these distributions exceed $\epsilon_{\rm jump2}=100\%$. Since the parental links built in Sect.~\ref{s:network} are constructed under the assumption that at least $75\%$ of a child area overlaps with its parent area (see Sect.~\ref{s:family}), flux and thus mass efficiency values greater than one can exist. These cases may correspond to situations in which the gas reservoirs of children extend beyond those of their immediate parental structures, as in the clump-fed accretion scenario \citep[see, e.g.,][]{motte2018a, peretto2020}. 
The efficiency distributions of the MHD synthetic protocluster is fitted using a Gaussian with $\overline{\epsilon_{\rm jump2}} \simeq 60\%\pm25\%$. In the W43-MM1 protocluster, we could not properly characterize the mass transfer efficiency distribution. This is due to the small number of sources over a six-times-larger scale span and the reasons given above. The median value of the flux transfer efficiency in W43-MM1, measured as $\sim$$60\%$, is a lower limit since the W43-MM1 catalogs are not expected to be as complete as in numerical simulations and some fluxes measured at high-resolution could be optically thick, thus underestimating fluxes of children. Therefore, we hereafter assumed that the mass transfer efficiency is $\overline{\epsilon_{\rm jump2}} \ge 60\%\pm25\%$ in W43-MM1.

The 3D fractality measured in the W43-MM1 protocluster is $\mathcal{F_{\rm 3D}}\simeq 1.19\pm 0.10$ (see Sect.~\ref{s:fractality}). This means that, for every jump in physical scales of a factor of two, one finds two children in a parental structure one-fifth of the time. As shown in Sect.~\ref{s:family} and \cref{fig:param-definition}, the mass partition between the dominant (primary) child and the other (secondary) child can be estimated by the mass ratio between these two children. The median of the flux ratios measured in W43-MM1 is $\sim$1.95. If we ignore temperature variations between children of the same parental structure, the median mass ratio would then be $\overline{\gamma}\simeq1.95$. The mass partition of fragments in W43-MM1 could therefore be imbalanced, MP(Primary; Secondary)~$\simeq [\frac{2}{3} \,;\, \frac{1}{3}]$, and similar to those found in the synthetic protoclusters examined here (see \cref{tab:fragm-casc_measured}). 

In summary, we showed that, in W43-MM1 and for every two-fold decrease in physical scales, parental structures should host single children four-fifths of the time and two children the other one-fifth of the time. Although the measured fluxes are not directly proportional to the masses, the mass transfer efficiency is expected to be high, and the mass partition is likely imbalanced (see \cref{tab:fragm-casc_measured}).

\begin{figure}[htbp!]
    \centering
    \includegraphics[width=0.5\textwidth]{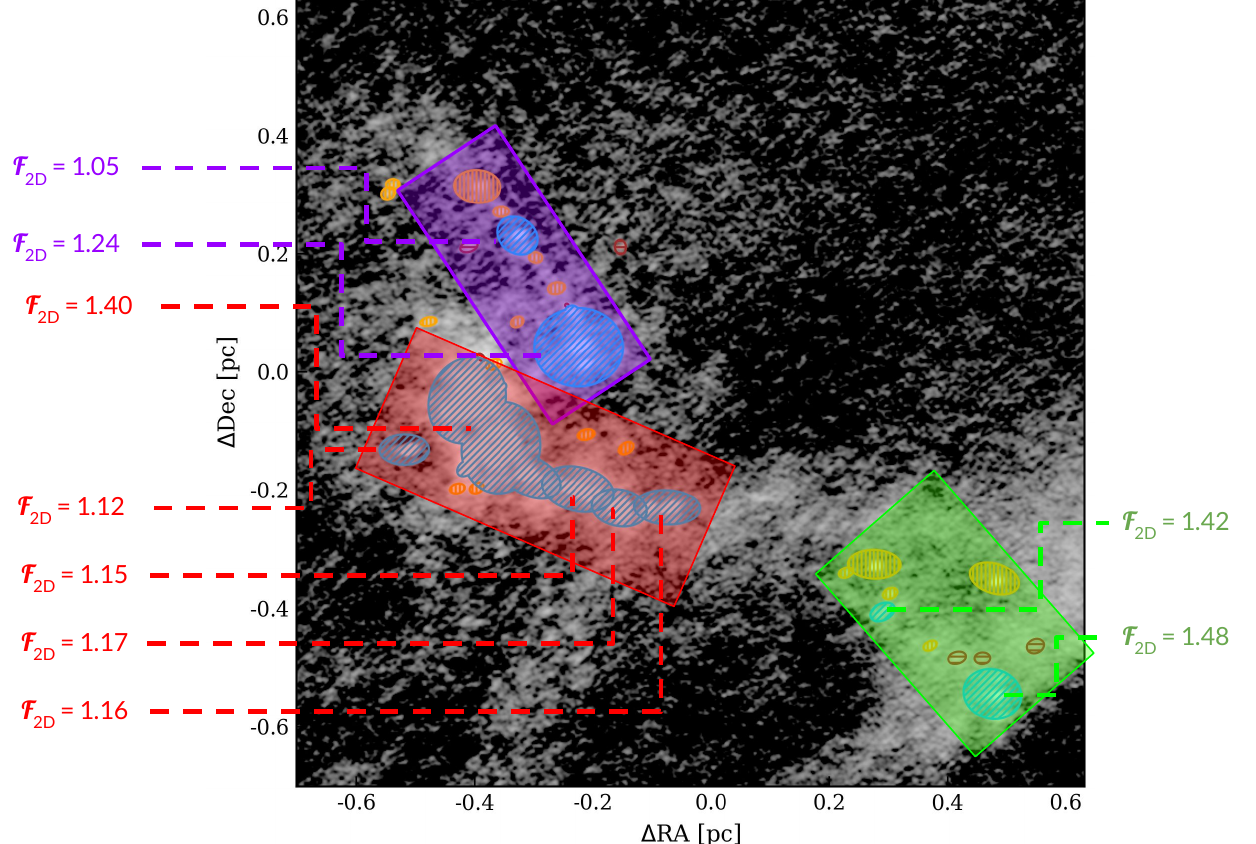}
    \caption{Spatial variation of the fractality index measured in the nine hierarchical structures of W43-MM1 (blue hatched areas), ranging from $\mathcal{F_{\rm 2D}}\simeq 1.0$ to $1.5$. A weak trend appears: the denser the gas (red and violet rectangular areas are denser than the green rectangular area), the richer the hierarchical structure and the smaller its fractality index. 
    }
    \label{fig:fractality-var}
\end{figure}
\section{Discussion}
\label{s:discussion}

In Sect.~\ref{s:results}, we characterized the mass cascade developing in the W43-MM1 protocluster. \cref{s:cascade-var} searches for variations of the cascade parameters in space and over scales. Then Sect.~\ref{s:cascade-comp} compares the W43-MM1 cascade with those measured in other protoclusters. \cref{s:fragmRegime} also discusses the resulting constraints on fragmentation levels and core formation efficiency (CFE), as defined below, within the framework of values reported in the literature. \cref{s:imf} finally predicts the mass function of gas reservoirs for star formation in the W43-MM1 protocluster. This is done by simulating the transfer of gas mass from cores identified at 2400~au \citep{nony2023} to seeds 200~au in size.

\subsection{Mass cascade variations in space and across scales}
\label{s:cascade-var}

As already mentioned by \cite{thomasson2024}, the fractality is expected to vary with the environment (density, turbulence level, etc.). Figure~\ref{fig:fractality-var} shows that the fractality indices of the W43-MM1 structures range from $\mathcal{F}_{\rm 2D}=1.0$ to 1.5, with denser hierarchical structures tending to have smaller indices. In contrast, we found no conclusive correlation between the fractality index measured below the 2.4~kau scale of cores and the core mass. The correlation does not strengthen separating prestellar and protostellar cores. However, our core sample is not large enough for this lack of correlation to be statistically significant. Further studies are necessary to investigate how the fractality index varies with the cloud environment and to statistically confirm that denser clumps are less fragmented.

To investigate how the fractality depends on the scale range, we calculated the fractality of W43-MM1 by separating the three scales above and below the core scale of 2.4~kau, thus excluding two scales from the dataset originally used by \FAMILY (see \cref{tab:sample}). We measured a fractality index of $\mathcal{F}_{\rm 3D}\simeq1.17\pm 0.35$ at scales smaller than $\sim$2.4~kau, which is very close to the average fractality index of W43-MM1 given in \cref{tab:fragm-casc_measured}. This is due to the significantly greater number of sources identified at higher resolutions, which directly results from the increased number of resolution elements in these maps. In contrast, at scales larger than 2.4~kau, the fractality index increases up to $\mathcal{F}_{\rm 3D}\simeq1.53\pm 0.2$. As discussed below, such an increase at large scales is expected when entering a regime dominated by turbulence \citep{thomasson2024}. These initial results must be confirmed by studies with larger statistics.

As stated in Sect.~\ref{s:MassTransfer}, the variation of mass transfer efficiency cannot be investigated using our current observational database. In the synthetic images, there is a slight tendency for the mass transfer efficiency to increase when focusing on the densest sources. It also slightly increases when considering in \cref{appendixfig:param-measured}, the population of child sources and their direct parents identified at increasingly higher resolutions. Therefore, while there are a few hints of variations in the mass cascade parameters, no firm conclusions can be drawn yet.

\subsection{Comparison of mass cascades}
\label{s:cascade-comp}

This study is the second to examine the hierarchical structure and mass cascade of a protocluster in detail. \cite{thomasson2022} applied \FAMILY to the intermediate-mass protocluster NGC~2264, which was imaged at scales ranging from 26 to 1.4~kau (from 0.13 to 0.007~pc). These scales are $2-5$ times larger than those studied in Sect.~\ref{s:results}, $14-0.27$~kau (from 0.07 to 0.001~pc), for the W43-MM1 protocluster. Like for synthetic protoclusters, the fragmentation cascade of NGC~2264 traces the fragmentation down to the core scale according to the definition by, for example, \cite{motte2018a}. Characterizing the fragmentation cascades of both NGC~2264 and synthetic protoclusters could in principle enable the study of turbulent fragmentation. Studying the fragmentation cascade of W43-MM1, on the other hand, could allow us to investigate two fragmentation regimes: one mainly driven by turbulence and the other by gravity, respectively at scales larger and smaller than the core scale.

As shown in \cref{tab:fragm-casc_measured}, the W43-MM1 protocluster has twice and four times as many hierarchical structures, percentage-wise, as the synthetic MHD and NGC~2264 protoclusters (see \cref{tab:fragm-casc_measured}). Conversely, it has up to twice less isolated structures. This could be related to the fact that the scales probed by synthetic catalogs and NGC~2264 study are larger and for NGC~2264 measured over an area seven times larger. Another possible reason is that W43-MM1 is undergoing a starburst event, which results in an intense star formation activity concentrated in space and time \citep{motte2003, nguyen2013}. Consequently, W43-MM1 hosts a more concentrated star formation event than synthetic Orion-like protoclusters. This concentration is not the result of W43-MM1 being evolved as it is one of the youngest protoclusters in the ALMA-IMF survey \citep[Papers~I and XVII,][]{motte2022, galvan2024}. ALMA images of the W43-MM1 protocluster revealed structures only within the inner 1.6~pc-diameter area selected here to analyze its structure (see \cref{appendixfig:spatial-distrib}). As illustrated in the 3~mm ALMA-IMF image in \cite{ginsburg2022}, areas located 0.5 to 2~pc from the field center are essentially empty. This cannot be attributed to the low sensitivity of the observations, at least not at the $0.43\arcsec$ scale corresponding to the deep ALMA-IMF mosaic \citep{nony2023}.

The 3D fractality index of W43-MM1 is smaller than that measured in the synthetic MHD protocluster: $\mathcal{F}_{\rm 3D}=1.19\pm 0.10$ and  $\mathcal{F}_{\rm 3D}=1.47\pm 0.15$, respectively (see \cref{tab:fragm-casc_measured}). It is also smaller than the measurement made in the intermediate-mass protocluster NGC~2264, $\mathcal{F}_{\rm 3D}=1.70\pm 0.24$ \citep{thomasson2024}. To obtain similar values in W43-MM1, the analysis must focus on the largest scales, $\mathcal{F}_{\rm 3D}=1.53\pm 0.2$ at $2.4-14$~kau scales (see Sect.~\ref{s:cascade-var}), or the lowest density regions (green rectangular area in \cref{fig:fractality-var}). This may seem contradictory to the results of \cite{thomasson2022}, which suggest that most linear structures, therefore with $\mathcal{F}_{\rm 3D}=1$, are found in the lower-density outskirts of the cloud. As mentioned earlier, no sources are yet present in the outskirts of the W43-MM1 starburst protocluster. Therefore, the low fractality values we have measured in the densest regions of W43-MM1 must reflect a change in the physical processes, from gravo-turbulence to a combination of gravity, magnetic fields and stellar feedback. Like shown in numerical simulations as a whole, the smaller number of total sources in the MHD synthetic protocluster compared to the HD protocluster illustrates how the magnetic field organizes the interstellar medium by concentrating its gas in specific areas. However, the MHD simulations of Sect.~\ref{s:simu} are based on ideal MHD equations and cannot represent the cloud structuration at scales smaller than ~5 kau or predict the associated fractality index. It is unclear whether the small fractality index measured in W43-MM1, $\mathcal{F}_{\rm 3D}=1.17\pm 0.10$ at $0.27-2.4$~kau scales, reflects the suppression of fragmentation by the magnetic fields or tidal interactions. However, \cite{valeille2026} measured rather high values of the magnetic field strength in the plane of the sky, up to 50~mG on the $\sim$2.4~kau scale of the W43-MM1 cores.

The mass transfer efficiency of the MHD synthetic protocluster is high, $\overline{\epsilon_{\rm jump2}} \simeq 60\%\pm25\%$, and tends to increase with density or decreasing scale (see Sect.~\ref{s:MassTransfer}). If confirmed, this trend would be consistent with the quasi-linear dependence of star formation efficiency with gas density at $0.1-1$~pc scales \citep{louvet2014} and the correlation of CFE with clump density \citep{bontemps2010, palau2013} or accretion rate with core mass \citep{wells2024, morii2026}. Regarding the mass partition, MP(Primary; Secondary)~$\simeq [\frac{2}{3} \,;\, \frac{1}{3}]$, the measurement in W43-MM1 is consistent with those proposed by \cite{pouteau2022} and \cite{yoo2025}, on the basis of flux ratios measured in high-mass core systems of massive protoclusters. In contrast, mass equipartition has been reported for a more isolated, massive prestellar core \citep{sanhueza2025}. The W43-MM1 mass partition is also consistent with the results obtained for the synthetic protoclusters analyzed here (see \cref{tab:fragm-casc_measured}). Unbalanced mass partitions are predicted in competitive accretion models, some of which are based on tidal-lobe accretion rates \citep[e.g.,][]{clarkWhitworth2021}. In fact, competitive accretion is a process that should develop in massive protoclusters, like W43-MM1 and W51-IRS2, where star formation occurs in local bursts.

\subsection{Fragmentation levels, CFE, and fragmentation regime}
\label{s:fragmRegime}

The fractality index of \cref{tab:fragm-casc_measured} enables the computation of the fragmentation or multiplicity level of a parental structure, defined as the number of fragments it contains. Several studies of high-mass star-forming regions have focused on fragmentation from the clump to the core scales \citep[e.g.,][]{bontemps2010, sanhueza2019, elia2026, morii2024}. More recently, a couple of studies characterized the fragmentation from the core to the seed scales \citep{budaiev2024, yoo2025, luo2026}. To compare these studies with each other, as well as with our own study, we must rescale the number of fragments to a given jump in physical scales, as is done when computing a fractality index. Using \cref{eq:fractality} and the $\mathcal{F}_{\rm 3D}=1.19$ of W43-MM1, we estimate that, on average, a clump detected at a scale of $s_{\rm clump}=30$~kau will fragment into $n_{\rm core} \simeq 2.0$ cores of scale $s_{\rm core}=2$~kau. On the other hand, a core will fragment into $n_{\rm seed} \simeq 1.8$ seeds of scale $s_{\rm seed}=0.2$~kau. The $n_{\rm core}$ number is, within a factor of two, equal to the number of cores detected in the most massive clumps of these studies \citep{bontemps2010, palau2013, traficante2023}. In contrast, it is six times smaller than those measured in infrared-dark clouds \citep{sanhueza2019}. The number of predicted fragments is also generally consistent with measurements performed by \cite{budaiev2024}, \cite{yoo2025}, and \cite{luo2026}. However, \cite{yoo2025} found that one of his targeted protocluster has three times fewer fragments than predicted. Therefore, it is likely that there are spatial variations in the fragmentation levels of high-mass star-forming regions but current studies are neither robust nor statistically significant enough to reveal the physical processes behind them. Studies based on more than two scales are also required to be robust against detection incompleteness and be corrected for undetected sources in the hierarchical cascade (see Sect.~\ref{s:fractality}). 

Additionally, the mass transfer efficiency estimated in Sect.~\ref{s:fractality} for the W43-MM1 fragmentation cascade predicts the CFE within a given parental structure, defined as the ratio of the sum of fragment masses to the parental structure mass. According to \cref{eq:eff} and $\overline{\epsilon_{\rm jump2}} \simeq 60\%$, the CFE of 2~kau cores within 30~kau clumps is predicted to be $\rm \epsilon^{clump}_{cores} \simeq 14\%$. The published CFE of \cite{bontemps2010}, \cite{palau2013}, and \cite{morii2023} are in excellent agreement with this value. However, measurements over more different scales or environments are $\sim$4 times smaller or larger \citep{traficante2023, yoo2025, elia2026}. This preliminary result suggests that the mass transfer efficiency varies with cloud properties such as density \citep{csengeri2017a} and accretion flows \citep{wells2024, morii2025}. To provide more robust estimates of the mass transfer efficiency and its spatial and scale-dependent variation, future studies must consider dust emissivity variations and temperature gradients.

Compared to the Jeans instability predictions, the fragmentation level of the massive cores present in the W43-MM1 protocluster, $\sim$2 seeds within $10-100~\msun$ cores, is small. This low level of fragmentation has already been observed in other high-mass star-forming regions, from clump to core scales \citep{bontemps2010, palau2013, traficante2023}, and thus appears to persist down to the 270~au scales studied here \citep[see also][]{ yoo2025, luo2026}. These results prove that gravitational fragmentation according to the Jeans criterion is invalid and that mass cascades can create massive, so-called super-Jeans structures \citep{bontemps2010, wang2011, palau2015, yoo2025}. The low fragmentation level observed in W43-MM1, particularly in its densest regions, is similar to that observed in the W51 and Sgr~B2 protoclusters \citep{budaiev2024, yoo2025}. These results support the formation of high-mass stars in such dense environments, as proposed by the hierarchical collapse and clump-fed accretion scenarios \citep[e.g.,][]{motte2018a, vazquez2019}.

This interpretation is confirmed when comparing our estimates of the fractality index and mass transfer efficiency in W43-MM1 with the predictive model of \cite{thomasson2024}. This gravo-turbulent model simulates the mass cascade in a star-forming cloud. It assumes that gravitational instabilities originate from turbulent shocks and takes into account the thermodynamics of the resulting chain of gravitational collapses. The model of \cite{thomasson2024} uses fragmentation rates defined as $\Phi=\ln\mathcal{F}_{\rm 3D}/\ln 2$ and mass transfer rates defined as $\xi=\ln\overline{\epsilon_{\rm jump2}}/\ln 2$. The fractality indices of $\mathcal{F}_{\rm 3D}=1.17$ at $0.27-2.4$~kau scales and $\mathcal{F}_{\rm 3D}=1.53$ at $2.4-14$~kau scales correspond to $\Phi \simeq 0.23$ and $\Phi\simeq0.61$. And the mass transfer efficiency, taken to be $\overline{\epsilon_{\rm jump2}} \gtrsim 60\%$, corresponds to $\xi \gtrsim -0.74$. When an adiabatic equation of state is assumed, the diagram of $\Phi$ as a function of size scale and $\xi$ allows to distinguish the fragmentation regime associated with supersonic turbulence from the thermal fragmentation regime, in which turbulence is subsonic \citep[see Fig.~7 of][]{thomasson2024}. In this figure, the two points corresponding to the W43-MM1 mass cascade lie in the second regime, in which fragmentation is driven by gravity and cloud structures are primarily supported by thermal energy. Across this $0.27-14$~kau range of scales, structures should therefore be (velocity) coherent, that is, mostly devoid of turbulent support \citep[e.g.,][]{ballesteros-paredes2011}. In low-mass star-forming regions, the transition to coherence is reached at the $\sim$8~kau scale of cores \citep{pineda2010, pokhrel2018, chen2019}. In W43-MM1, however, this transition occurs at larger scales, contrary to expectations for high-mass protoclusters from the large line widths observed for cores \citep[][including Paper~VII]{wienen2012, kauffmann2013, cunningham2023}. This result confirms a recent study showing that the turbulent and magnetic supports of W43-MM1 cores are substantially overestimated due to line width contamination by organized motions such as gravitational infall \citep{valeille2026}, as predicted by \cite{vazquez2026}. Our results also suggest that cores and seeds with super-Jeans masses form from hierarchical infalling gas flows that are coupled with magnetic fields and rotation. These flows could channel mass at a few seed locations while dispersing lower-density fluctuations through tidal effects.

\begin{figure}[htbp!]
    \centering
    \includegraphics[width=0.485
    \textwidth]{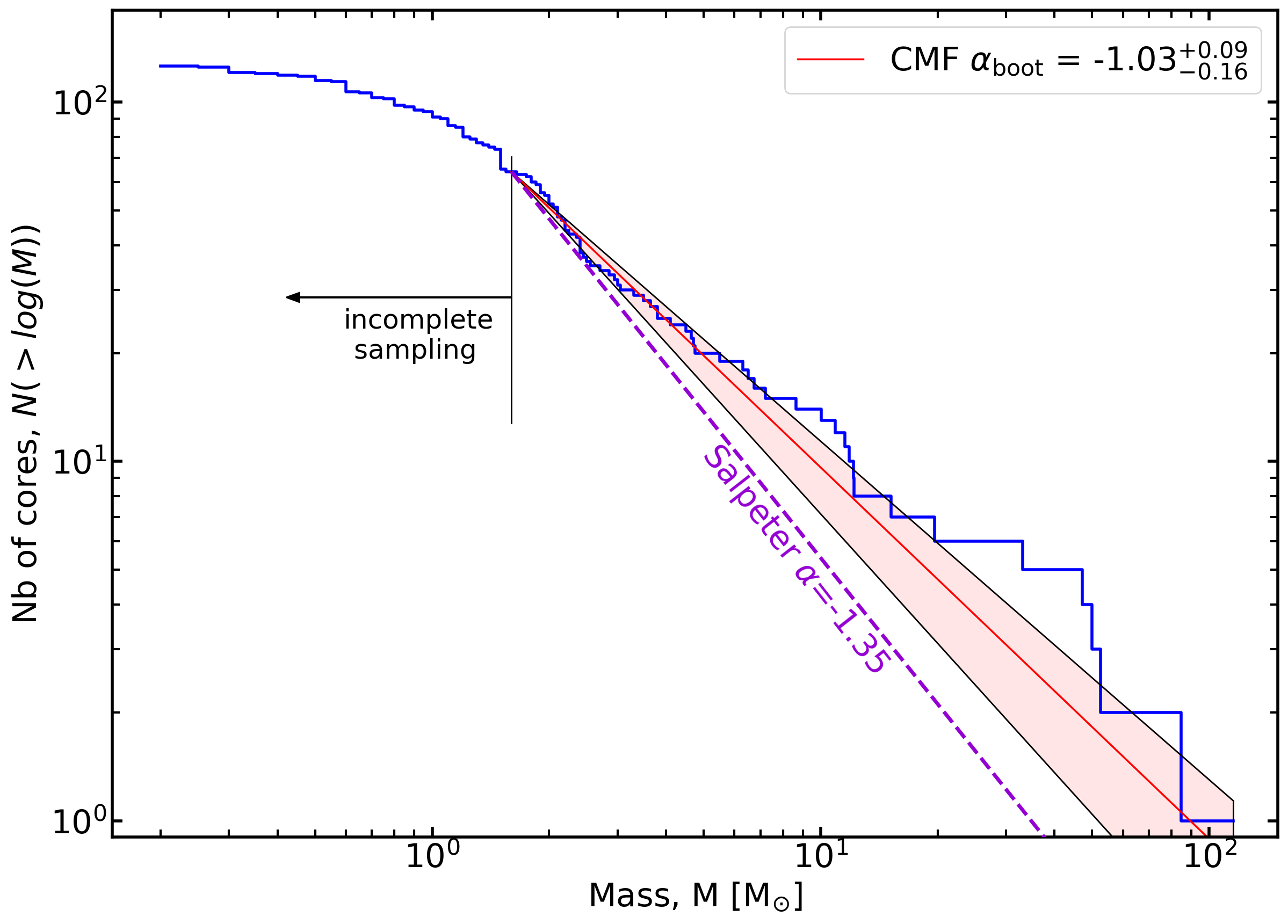}
    \vskip 0.2cm \includegraphics[width=0.485\textwidth]{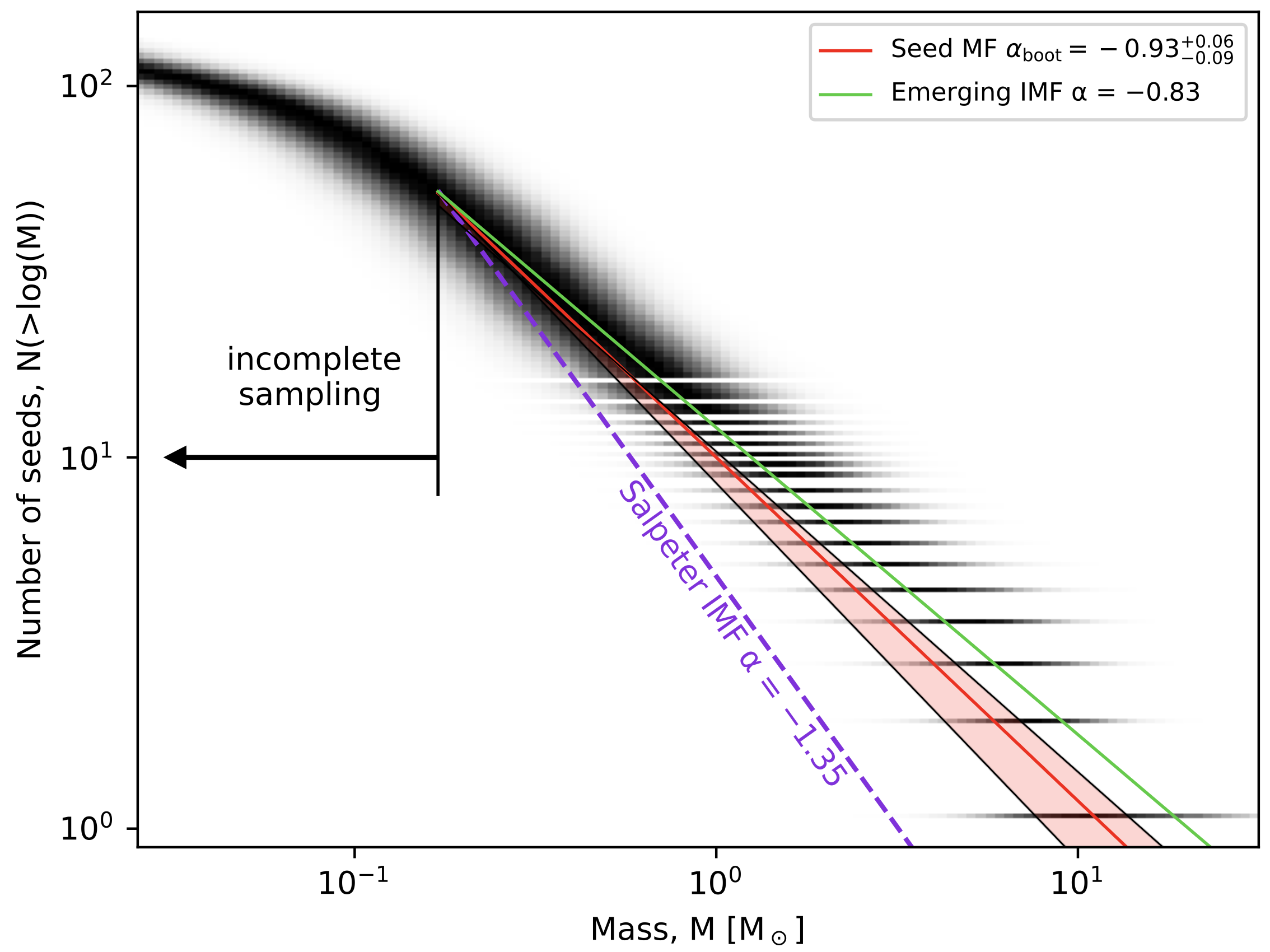}
    \vskip -0.1cm
    \caption{Mass functions of cores and seeds in W43-MM1, along with the emerging IMF that we predict here. \textit{Top panel:} Cumulative CMF of cores (blue histogram) extracted by \cite{nony2023}, with masses estimated by \cite{motte2025}. \textit{Bottom panel:} Cumulative mass function (MF) of seeds, which are randomly drawn (gray point density) from the mass cascade of Sect.~\ref{s:results} applied to the CMF. The emerging IMF (green line) is predicted from the seed MF, with seeds fed proportionally to their mass (see Sect.~\ref{s:imf}). The high-mass ends of these three mass functions are described by $\frac{dN}{d\log M} \propto M^\alpha$ relationships. The CMF and seed MF have slopes (red lines) and $1\sigma$ uncertainties resulting from the \cite{alstott2014} and bootstrap methods (pink areas). All of these functions are flatter that the Salpeter slope of the canonical IMF (dashed magenta lines, $\alpha\simeq -1.35$): $\alpha\simeq -1.03^{+0.09}_{-0.16}$ for the CMF, $\alpha\simeq-0.92^{+0.06}_{-0.09}$ for the seed MF, and $\alpha\simeq-0.83^{+0.06}_{-0.09}$ for the emerging IMF.}
    \label{fig:cmf}
\end{figure}

\subsection{Predicted emerging IMF}
\label{s:imf}

Here, we determine and \cref{fig:cmf} presents the emerging IMF, which we assume to be equivalent to the mass distribution of the gas reservoir that will be used to form individual stars in W43-MM1. To account for core subfragmentation, we start from the CMF of the W43-MM1 cores and predict the mass function of seeds with a characteristic size of 200 au. The 127 W43-MM1 cores were extracted from the best-sensitivity denoised image of the ALMA-IMF Large Program at 1.3~mm  \citep[see][]{nony2023}. Their masses and uncertainties were updated by \cite{motte2025} using carefully determined dust temperatures. In agreement with the location of W43-MM1 protostars in the mass-luminosity diagram of \cite{motte2025}, we here assume that the core masses dominate the mass of their protostellar embryos. Above the completeness level of $\sim$$1.6~\msun$ \citep{motte2018b}, we fitted the CMF high-mass end by a single power-law of the form $\frac{dN}{d\log M} \propto M^\alpha$. Following previous ALMA-IMF studies \citep{pouteau2023, nony2023, armante2024, louvet2024}, we used a bootstrap procedure and the \textsl{powerlaw} python package of \cite{alstott2014} based on the maximum likelihood estimator method of \cite{clauset2009}. Doing so, we measured a CMF high-mass end slope of $\alpha=-1.03^{+0.09}_{-0.16}$ (see \cref{fig:cmf}a), which is consistent with the values reported in \cite{motte2018b} and \cite{nony2023}, and which confirms its top-heavy nature.

We then simulated a fragmentation cascade starting at the core scale and ending at the disk scale, that is, from 2400~au to 200~au for stars with $1-100~\msun$ masses (see \cref{fig:param-definition}). Below the core scale, the model of \cite{thomasson2024} predicts that the mass cascade parameters remain constant until the disk scale is reached, at which point rotation becomes dominant. The disk size is computed considering the dependence of disk size on the stellar mass, $R_{\rm disk} \propto M_\star^{1/3}$ \citep[see Eq.~13 of][]{hennebelle2016b}, and assuming 50~au for disk around  solar-type protostars \citep{maury2019, tobin2020, Yen+Li2024}. Unlike in previous studies \citep{pouteau2022, zhou2025, thomasson2026}, our prediction uses the data-informed parameters of the hierarchical mass cascade in W43-MM1. In detail, the $\mathcal{F}_{\rm 3D}=1.19\pm0.10$ fractality index, $\overline{\epsilon_{\rm jump2}} = 60\%\pm25\%$ mass transfer efficiency, and MP(Primary; Secondary)~$\simeq [\frac{2}{3} \,;\, \frac{1}{3}]$ mass partition derived in Sect.~\ref{s:results} are used in the model by \cite{thomasson2026}. Based on a combined stochastic and gravo-turbulent approach, this model introduces a framework for tracking the hierarchical fragmentation of a cloud. Taking into account the $\sim$10\% of the 3~mm cores that disappear at high resolution and the uncertainties in core masses, we generated $100\,000$ datasets of 2400~au cores from a random draw with discount of the W43-MM1 core sample presented above. According to \cref{eq:fractality} and given the measured $\mathcal{F}_{\rm 3D}=1.19$ (see Sect.~\ref{s:fractality}), a W43-MM1 core detected at a scale of $s_{\rm core}=2400$~au will fragment into $n_{\rm seed}\simeq1.9$ seeds of scale $s_{\rm seed}=200$~au. According to \cref{eq:eff} and with $\overline{\epsilon_{\rm jump2}} = 60\%$ (see Sect.~\ref{s:MassTransfer}), the efficiency of the formation of seeds is $\epsilon^{\rm2400~au}_{\rm 200~au}\simeq$16\%. Therefore, the completeness limit multiplied by this efficiency and the proportion of mass captured by the primary seed (see Sect.~\ref{s:MassTransfer}) becomes the completeness limit of the seed mass function: $16\%\times \frac{2}{3}\times 1.6~\msun \simeq 0.17~\msun$. 

Figure~\ref{fig:cmf}b displays the seed mass function of the W43-MM1 protocluster derived from the point density of these datasets generated by random draw. Above the completeness level of $\sim$$0.17~\msun$, their high-mass end is fitted with the same procedure as for the CMF by a power-law of the form $\frac{dN}{d\log M} \propto M^\alpha$ with $\alpha=-0.92^{+0.06}_{-0.09}$. This power-law remains flatter than the canonical IMF slope of \cite{salpeter1955}. We therefore have proven, on real data, the behavior found by \cite{thomasson2026}, which states that the power-law slope of the CMF high-mass end remains unchanged unless the fragmentation cascade parameters depend on the mass of cores. In Sect.~\ref{s:cascade-var}, we revealed some spatial and scale-dependent variations in the fragmentation cascade, but the present study cannot determine variation recipes. 
In a first attempt to move beyond these homogeneous fragmentation parameters and account for core mass growth, we  assumed that the masses of the star formation reservoirs correspond to the masses of their seeds at a $\sim$200~au scale once they had grown. We chose a mass growth rate proportional to the core mass, reminding the CFE and mass accretion rate dependence with density and mass, respectively found by \cite{louvet2014} and \cite{morii2025}. We applied a relation of $M_{\rm reservoir} = M_{\rm seed} \times (1+ \mu)$, with $\mu= \frac{M_{\rm seed}}{5~M_\odot}$ set for the largest mass reservoir of W43-MM1 to be $100~\msun$, to predict the mass function of the gas reservoirs that could be seen as the emerging IMF. The power-law fitted above the new completeness limit of $0.18~\msun$ becomes even flatter: $\frac{dN}{d\log M} \propto M^{-0.83}$. 

At this stage, the data-informed parameters of the fragmentation cascade investigated in W43-MM1 shows that the IMF emerging from its top-heavy CMF will most probably remain top-heavy. Stellar feedbacks or stellar dynamics could however change this picture.

\section{Conclusion} \label{s:conc}

We have characterized the fragmentation cascade of the W43-MM1 protocluster and investigated the impact of core sufragmentation on its CMF, which was found to be top-heavy \citep{motte2018b, nony2023}. Our methodology and main findings are summarized below:
\begin{itemize}
    \item Using \getsf, we extracted compact sources in a set of five ALMA images of the W43-MM1 protocluster, covering the $0.27-14$~kau resolution range (see Sect.~\ref{s:data-cat}, Figs.~\ref{fig:input-images-Main} and \ref{appendixfig:input-images-SW}, and Tables~\ref{tab:sample} and \ref{tab:MM1sources}). For comparison, we used source catalogs from a synthetic Orion-like protocluster derived from MHD numerical simulations (see \cref{tab:sample}).
    \item We employed \FAMILY \citep[][also described in Sect.~\ref{s:family}]{thomasson2022, thomasson2024}, a tool defining networks of nested sources in protoclusters (see Sect.~\ref{s:network}, Figs.~\ref{fig:struct-types}--\ref{fig:MM1-ExampleHierarchical} and \ref{appendixfig:Hierarchical}--\ref{appendixfig:spatial-distrib}). In Sects.~\ref{s:fractality}--\ref{s:MassTransfer}, we characterized the fragmentation cascades of W43-MM1 and synthetic protoclusters, which are at first order assumed to be scale-free. 
    \item The resulting cascade parameters are the fractality index, mass transfer efficiency, and mass partition between siblings defined in Sect.~\ref{s:family} (see also Figs.~\ref{fig:param-definition}, \ref{fig:fractality-missing}, and \ref{appendixfig:3Dfractality}--\ref{appendixfig:param-measured}). In W43-MM1, we found a low fractality index and imbalanced mass partition (see \cref{tab:fragm-casc_measured}): $\mathcal{F}_{\rm 3D}\simeq 1.19\pm 0.10$ and MP(Primary; Secondary)~$\simeq [\frac{2}{3} \,;\, \frac{1}{3}]$. We also estimated a high mass transfer efficiency of $\overline{\epsilon_{\rm jump2}} \gtrsim 60\%$.
    \item Mass cascade parameters are expected, but not yet firmly observed, to vary in space, scale, and during the concentrated star-formation event of the W43-MM1 mini-starburst (see Sect.~\ref{s:cascade-var}, Figs.~\ref{fig:fractality-var} and \ref{appendixfig:spatial-distrib}). A comparison with the fractality indices measured, only at large scales, in the NGC~2264 intermediate-mass and Orion-like synthetic protoclusters (see Sect.~\ref{s:cascade-comp}) shows that in W43-MM1 the fractal index, and thus fragmentation level, decreases at scales smaller than $\sim$2.4~kau: from $\mathcal{F}_{\rm 3D}=1.53$ to $\mathcal{F}_{\rm 3D}=1.17$ (see Sects.~\ref{s:cascade-var} and \ref{s:fragmRegime}).
    \item The mass cascade developing in W43-MM1 suggests, according to the predictive model of \cite{thomasson2024}, that its structures up to at least 14~kau lie in the regime of fragmentation driven by gravity (see Sect.~\ref{s:fragmRegime}). 
    \item We used the parameters of the mass cascade measured for W43-MM1 (see \cref{tab:fragm-casc_measured}) and the model by \cite{thomasson2026} to predict the subfragmentation of its 2.4~kau cores. The resulting seed mass function, from which the IMF will emerge, has a high-mass end which remains top-heavy (see Sect.~\ref{s:imf} and \cref{fig:cmf}). Therefore, based on our current assumptions, core subfragmentation has a minimal role in shaping the IMF.
\end{itemize}

Combining the characterization of the mass cascade in a protocluster with the \cite{thomasson2024} model is a powerful, robust method for determining the fragmentation regime of its cloud structures and the main physical processes at their origin. This information is invaluable, as line and polarization observations struggle to measure non-thermal support in massive protoclusters \citep[see, e.g.,][]{valeille2026}. 

Furthermore, statistical studies of mass cascades are mandatory to investigate how the cascade varies with cloud properties and evolutionary state. This is especially important for determining the shape of the mass function of seeds emerging from core subfragmentation. Additionally, investigation of the direct link between gas reservoirs devoted to star formation and stars requires constraining the growth of seed mass during the protostellar collapse. In fact, a tight connection between the structure cascade of clouds and the accretion cascade of infalling gas toward newborn stars is expected in the hierarchical collapse model \citep[e.g.,][]{vazquez2019}.

\section*{Data Availability}
The complete \cref{tab:MM1sources} is available in electronic form at the CDS via anonymous ftp to \href{cdsarc.u-strasbg.fr}{cdsarc.u-strasbg.fr} (130.79.128.5) or via \href{http://cdsweb.u-strasbg.fr/cgi-bin/qcat?J/A+A/}{http://cdsweb.u-strasbg.fr/cgi-bin/qcat?J/A+A/}. 


\begin{acknowledgements}
We thank the anonymous referee for strengthening the paper. This paper makes use of the following ALMA data: ADS/JAO.ALMA\#2017.1.01355.L, \#2018.1.01787.S, and \#2019.1.00502.S. ALMA is a partnership of ESO (representing its member states), NSF (USA) and NINS (Japan), together with NRC (Canada), MOST and ASIAA (Taiwan), and KASI (Republic of Korea), in cooperation with the Republic of Chile. 
This project has received funding from the European Research Council (ERC) via the ERC Synergy Grant \textsl{ECOGAL} (grant 855130) and from the French Agence Nationale de la Recherche (ANR) through the project \textsl{COSMHIC} (ANR-20-CE31-0009).
AG acknowledges the support of the PCMI of CNRS/INSU with INC/INP co-funded by CEA and CNES.
AS and NCT gratefully acknowledge support by the Fondecyt Regular (project code 1220610), and ANID BASAL project FB210003. AS is gratefully supported by the China-Chile Joint Research Fund (CCJRF No. 2312). 
MG acknowledges funding from MICIU/AEI/10.13039/501100011033 and FEDER, EU through project TACOS (PID2023-146635NA-I00).
RGM acknowledges the support from UNAM-PAPIIT project IN105225.
RAG acknowledges support from the STFC (grant ST/Y002229/1).
MB has received financial support from the NSF Astronomy \& Astrophysics program (grant number 2510129).
NSG gratefully acknowledges support from ANID Beca Doctorado Nacional 21250244. 
PS was partially supported by a Grant-in-Aid for Scientific Research (KAKENHI Number JP23H01221) of JSPS. 
\end{acknowledgements}

\bibliographystyle{aa}  
\bibliography{ALMA-IMF-Motte2}

\begin{appendix}


\section{Complementary figures}
\label{appendixsect:figures}

\begin{figure}[htbp!]
    \centering
    \includegraphics[width=1\textwidth]{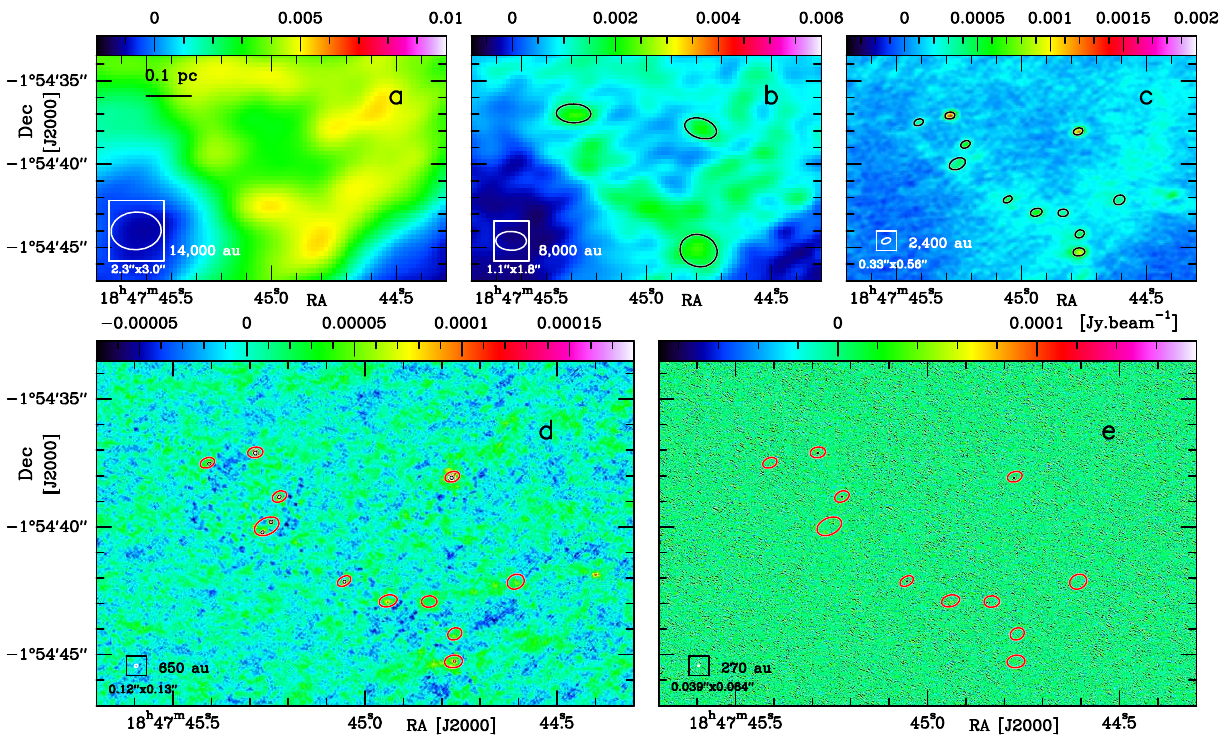} 
    \vskip -0.1cm
    \parbox{\textwidth}{\caption{South-western part of the W43-MM1 protocluster, as revealed by five 3~mm continuum ALMA images. 
    Same caption as \cref{fig:input-images-Main}. In \textit{panels d--e}, red ellipses outline sources extracted at a resolution of 2400~au.}}
    \label{appendixfig:input-images-SW}
\end{figure}

\begin{figure*}[htbp!]
    \centering
    \vskip -0.4cm   
    \includegraphics[width=0.9\textwidth]{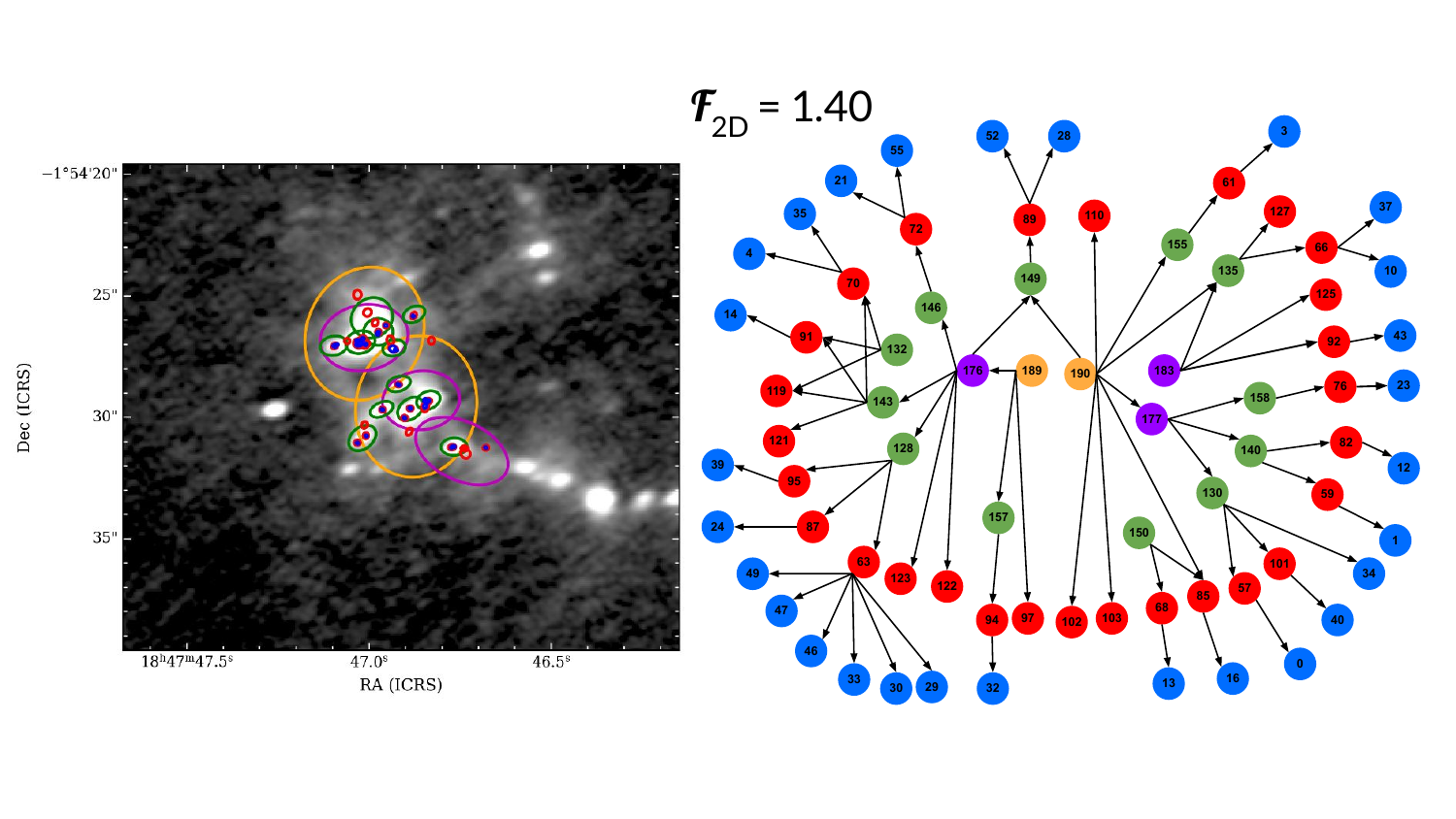}
    \vskip -0.9cm
    \includegraphics[width=0.51\textwidth]{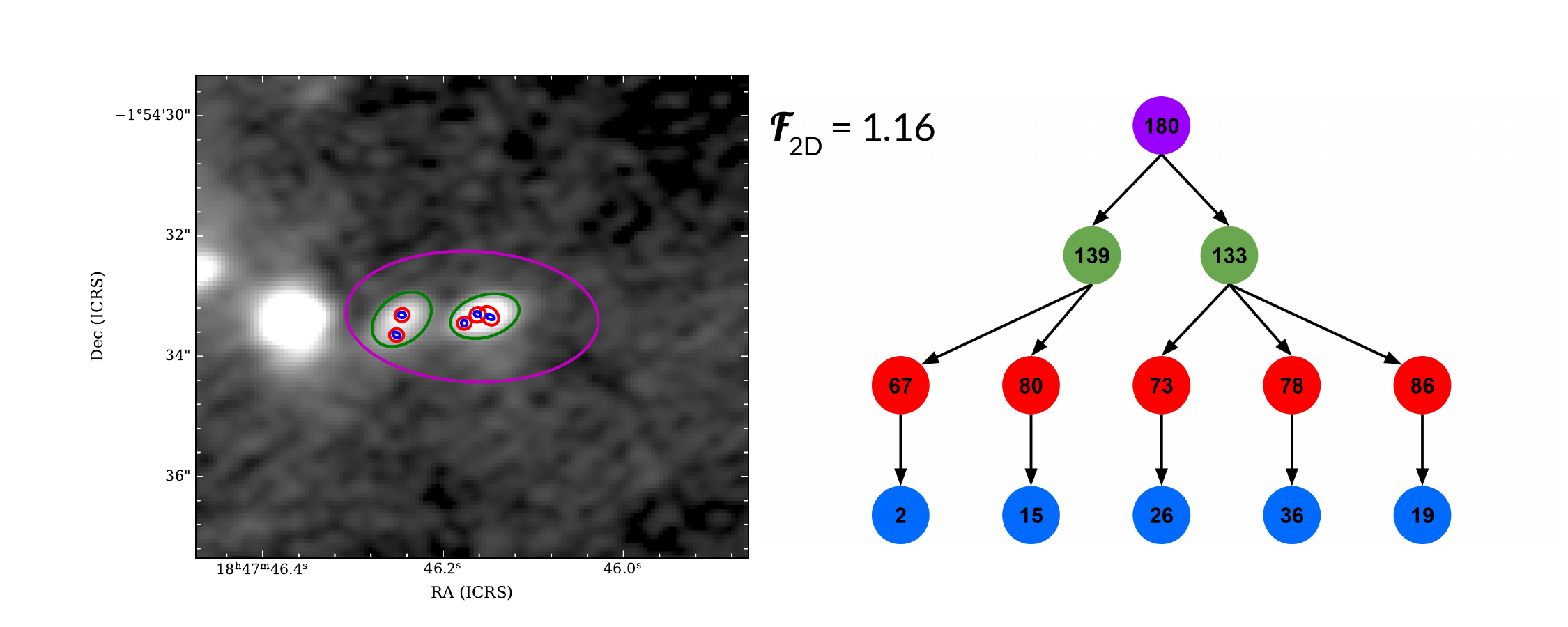}\hskip -0.7cm
    \includegraphics[width=0.51\textwidth]{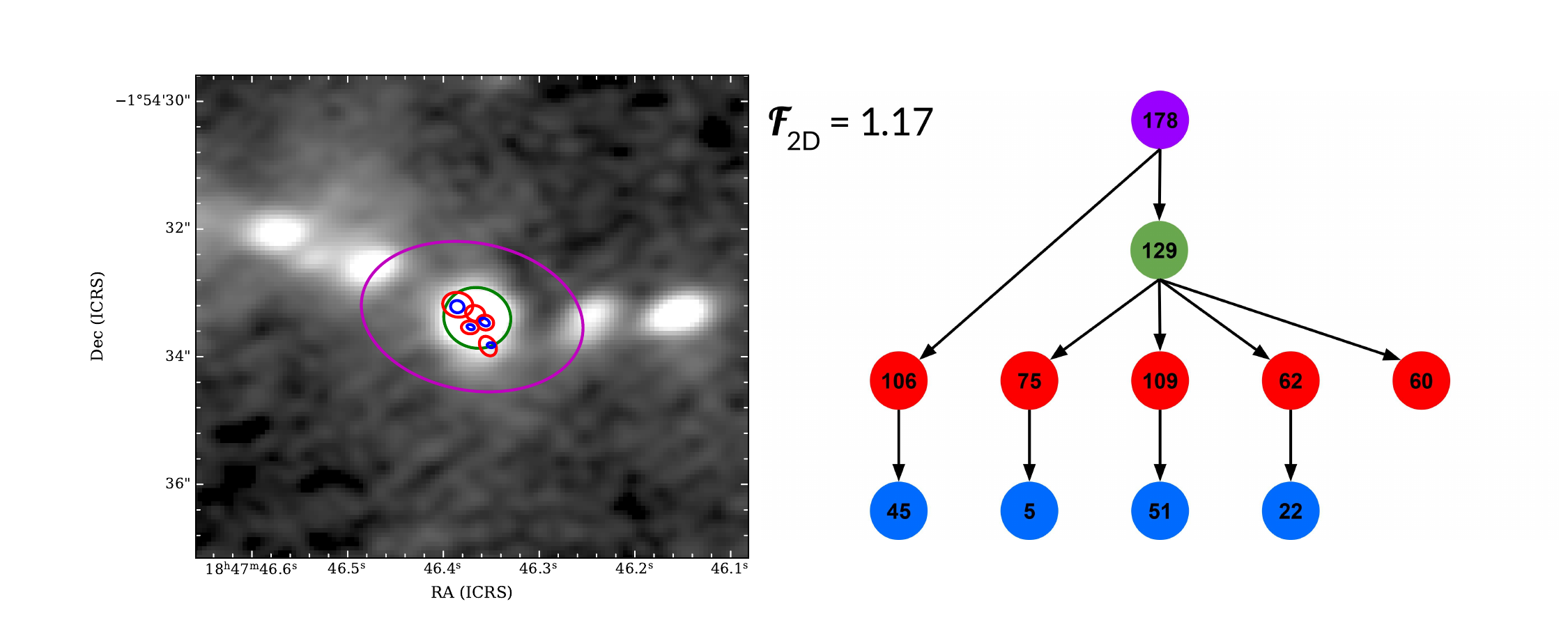}
    \includegraphics[width=0.51\textwidth]{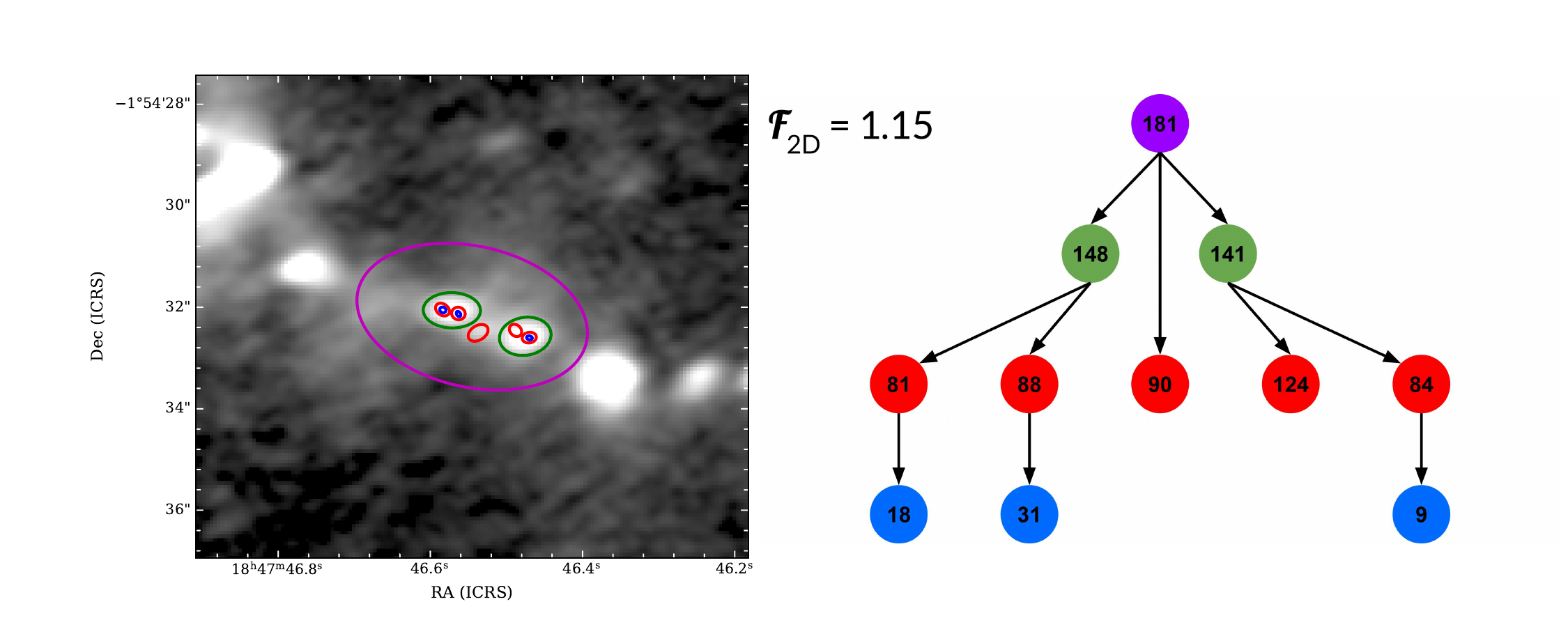} \hskip -0.8cm
    \includegraphics[width=0.51\textwidth]{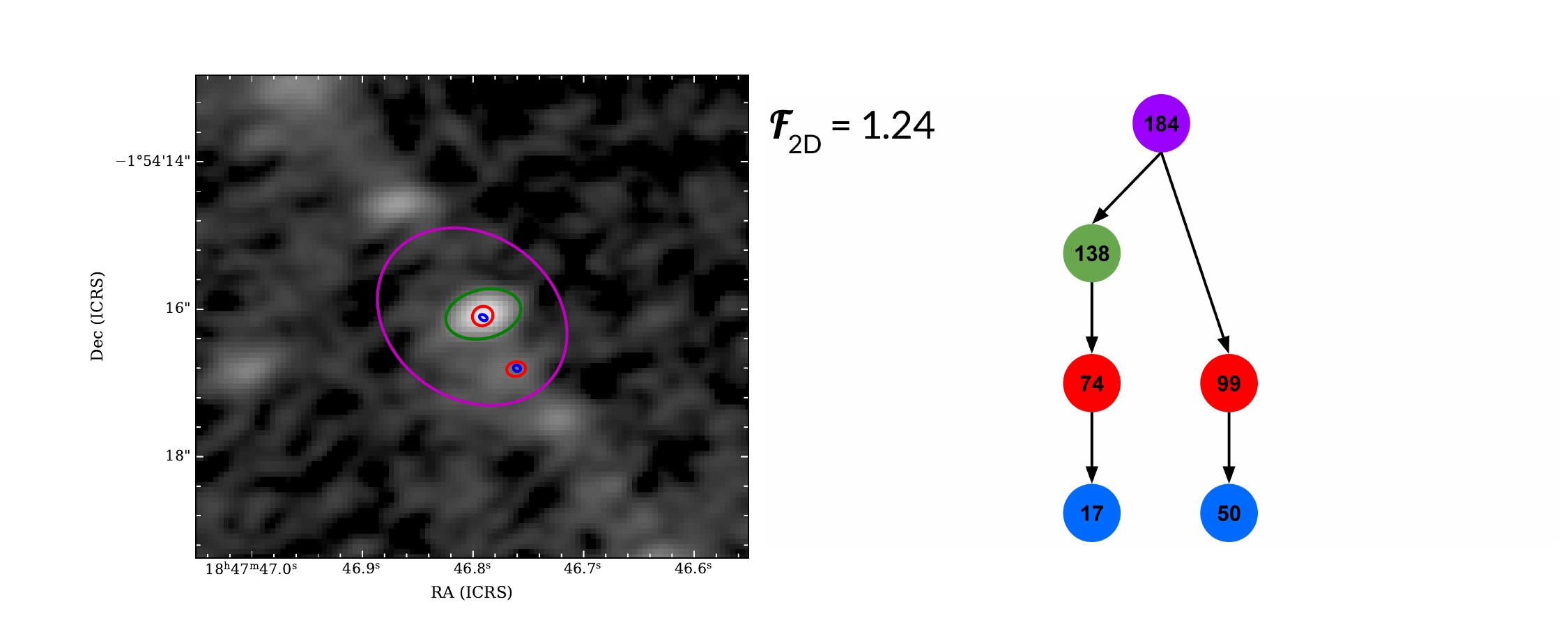}
    \includegraphics[width=0.51\textwidth]{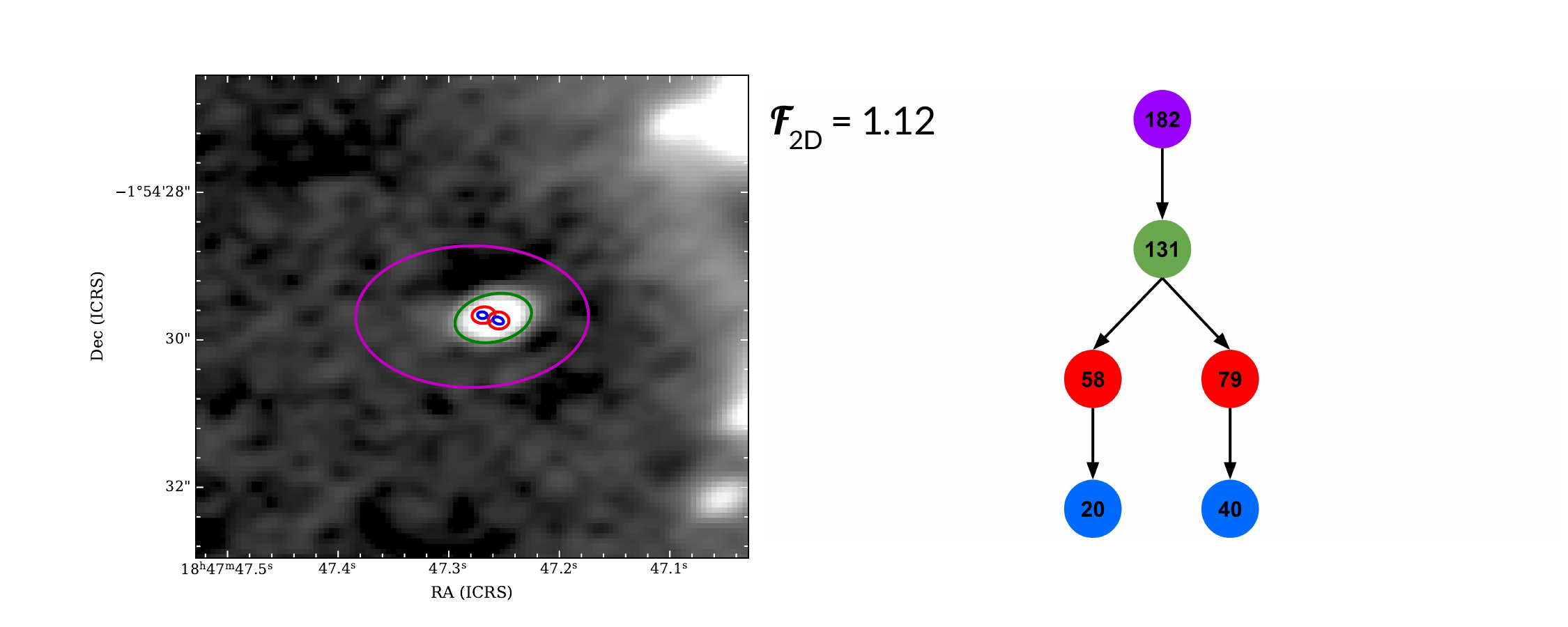} \hskip -0.9cm
    \includegraphics[width=0.51\textwidth]{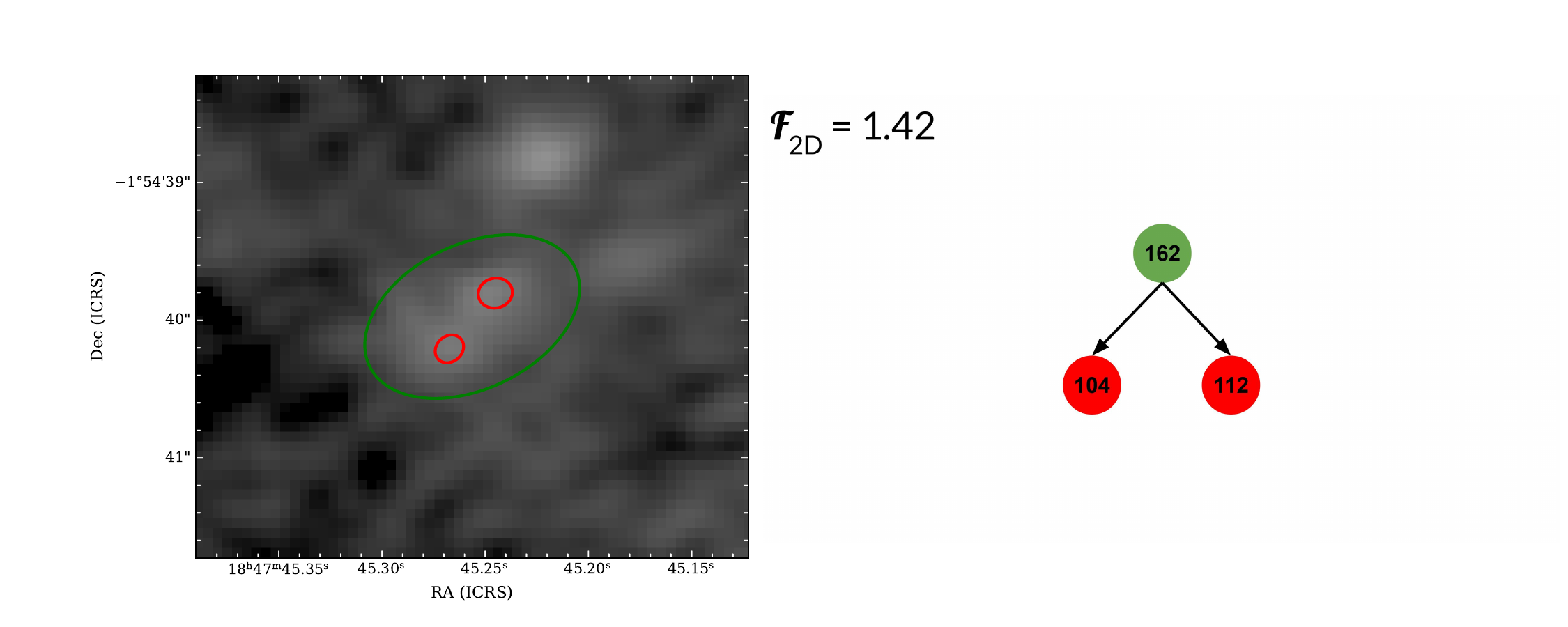}
    \includegraphics[width=0.52\textwidth]{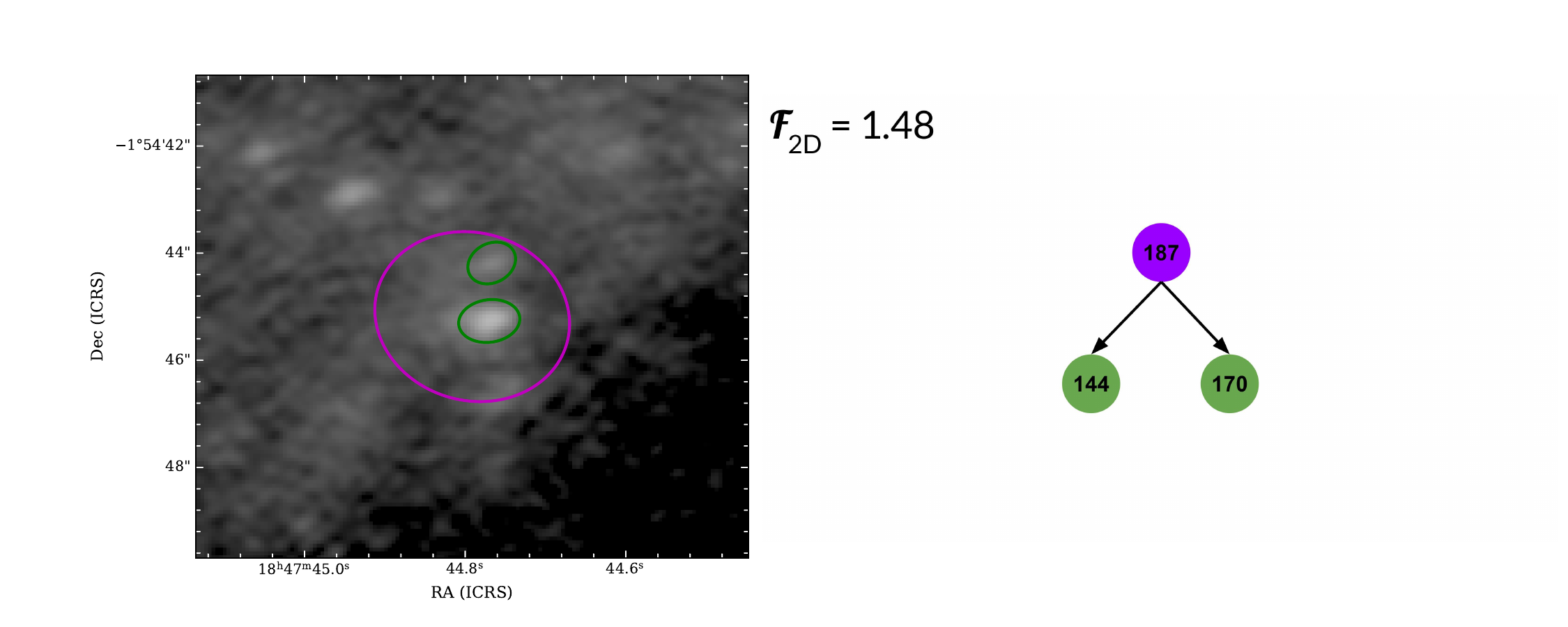}
    \caption{Hierarchical structures of the W43-MM1 protocluster, revealed by \textsl{FAMILY}. 
    \textit{Left panels:} Stack sources identified at resolutions of 14~kau (yellow), 8~kau (purple), 2.4~kau (green), 650~au (red), and 270~au (blue). The underlying image represents the 3~mm continuum imag at $0.43\arcsec$ (or 2400~au) resolution. Ellipses here represent the outer diameter of the sources, which is $\sim$1.7 times the FWHM of the ellipses in Figs.~\ref{fig:input-images-Main} and \ref{appendixfig:input-images-SW} (see \cref{s:intro}). 
    \textit{Right panels:} Representation of the associated hierarchical structure, showing the relationship established between sources of each scale, here identified by their node number (see also \cref{tab:MM1sources}).}
    \label{appendixfig:Hierarchical}
\end{figure*}

Figures~\ref{appendixfig:input-images-SW} and \ref{appendixfig:Hierarchical} present the five ALMA images covering the south-western part of W43-MM1 and the hierarchical structures characterized in the W43-MM1 protocluster, respectively complementing Figs.~\ref{fig:input-images-Main} and \ref{fig:MM1-ExampleHierarchical}. Then, the spatial distributions of the structures identified in the W43-MM1 and MHD synthetic protoclusters are illustrated in \cref{appendixfig:spatial-distrib}. Finally, \cref{appendixfig:3Dfractality} presents the 3D fractality indices predicted from their measured 2D analogs and \cref{appendixfig:param-measured} displays the distribution of mass transfer efficiencies measured for synthetic simulations.

\begin{figure}[htbp!]
    \hskip -1cm \includegraphics[width=0.55\textwidth]{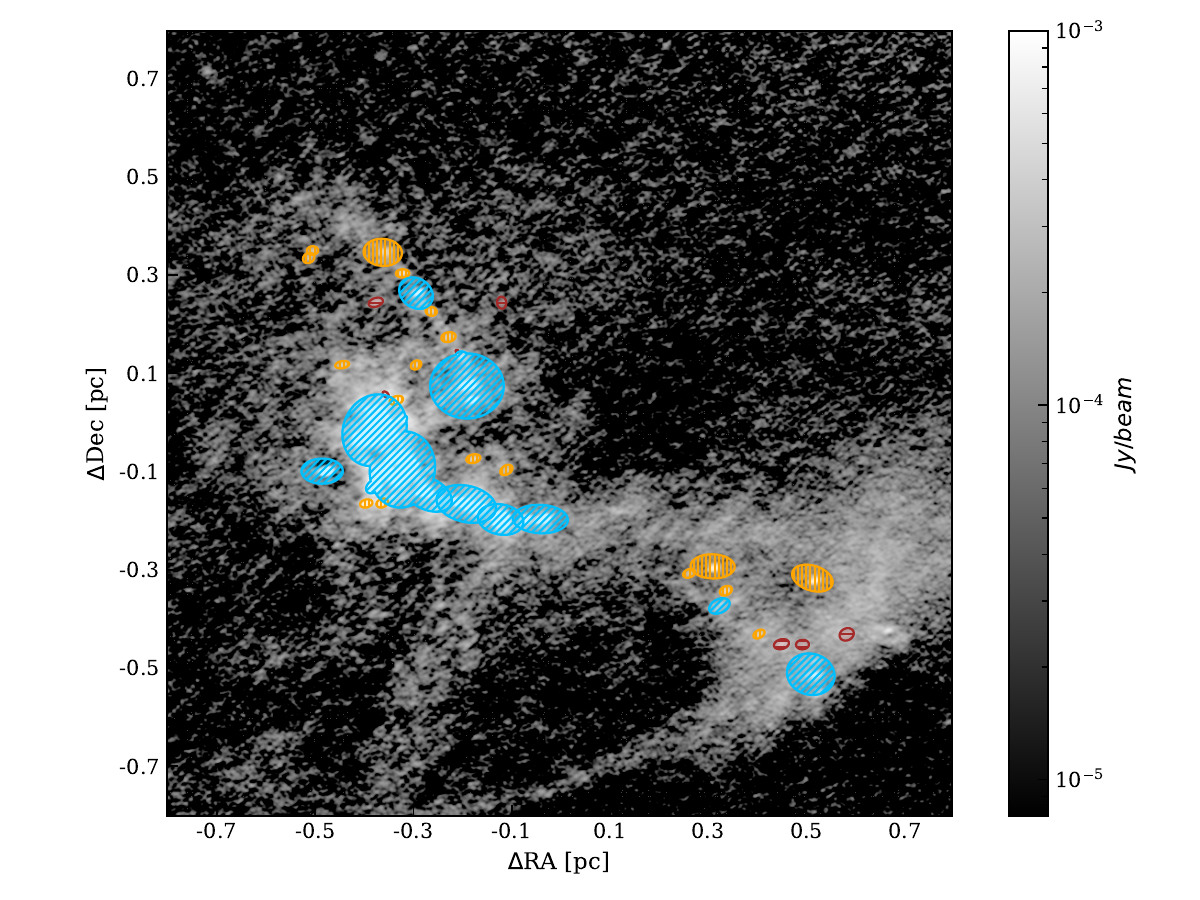}
    \vskip -0.25cm 
    \hskip -1.2cm \includegraphics[width=0.60\textwidth]{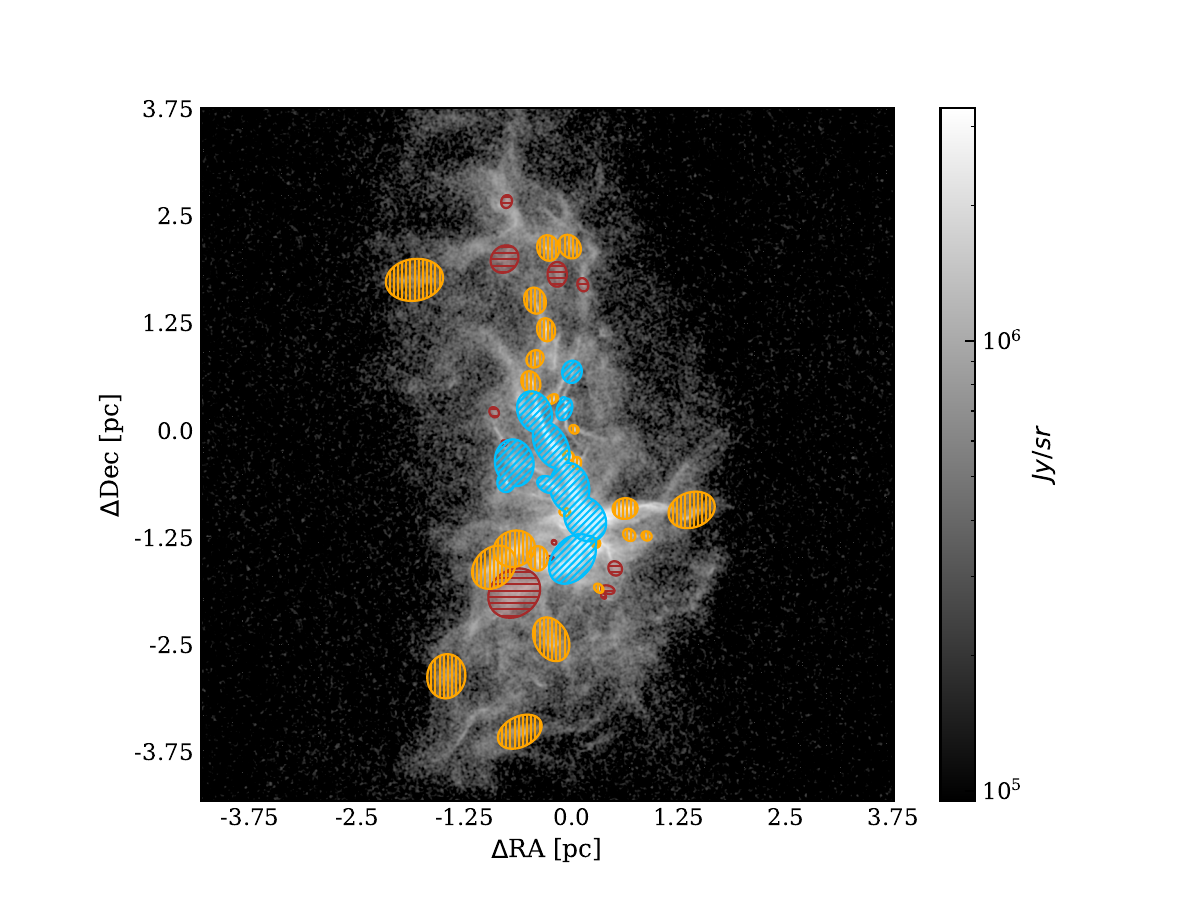}
    \vskip -0.3cm 
    \caption{Spatial distribution of structures identified in the W43-MM1 (\textit{top panel}) and MHD synthetic protoclusters (\textit{bottom panel}). Cyan, orange, and brown hashed ellipses outline hierarchical, linear, and isolated structures, respectively. W43-MM1 presents a dense cluster of hierarchical structures toward the ridge/hub. The background images (gray scale) are the 3~mm intensities at spatial resolutions of 2.4~kau in \textit{top panel} and 5~kau in \textit{bottom panel}, respectively.}
    \label{appendixfig:spatial-distrib}
\end{figure}

\FloatBarrier
\begin{figure}[htbp!]
    \centering
    \vskip 0.2cm
    \includegraphics[width=0.5\textwidth]{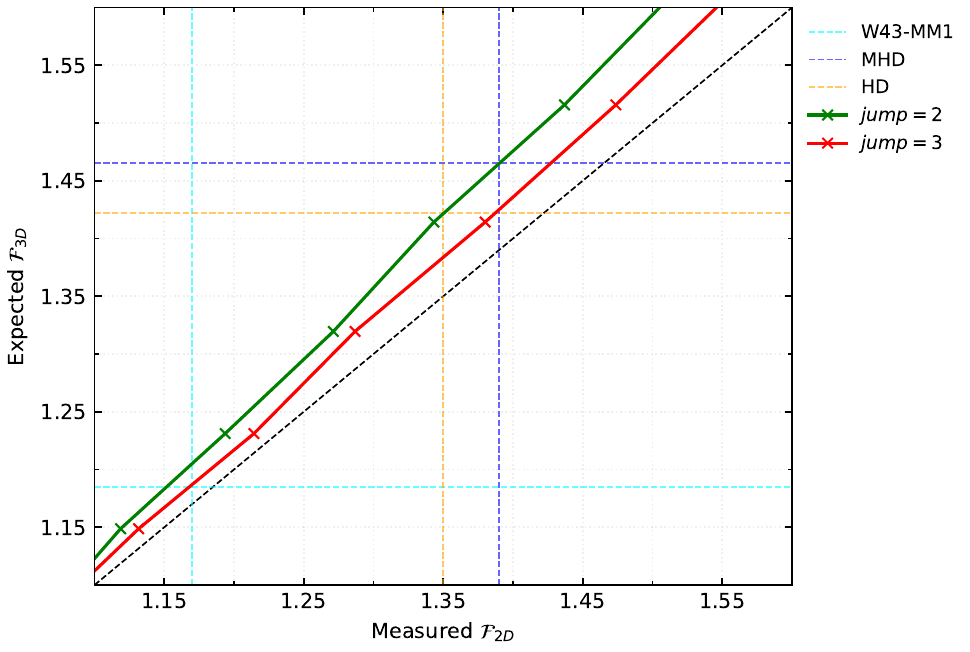}
    \vskip -0.0cm
    \caption{The 3D fractality index, $\mathcal{F}_{\rm 3D}$, predicted as a function of the measured 2D fractality index, $\mathcal{F}_{\rm 2D}$, here compared to the unity line (black dashed). Predictions were computed for jumps in physical scales of $\rm \textit{jump}=2$ (green curve) and $\rm \textit{jump}=3$ (red curve), which correspond to the synthetic and W43-MM1 protocluster datasets, respectively (see \cref{tab:sample}). The values of the 3D and 2D fractality indices plotted here are, in the present study, rescaled to jumps in physical scales of two (see Sect.~\ref{s:family}). Projection effects are smaller for the W43-MM1 protocluster than for the synthetic protoclusters.}
    \label{appendixfig:3Dfractality}
\end{figure}

\begin{figure}[htbp!]
    \centering
    \vskip 0.1cm
    \includegraphics[width=0.235\textwidth]{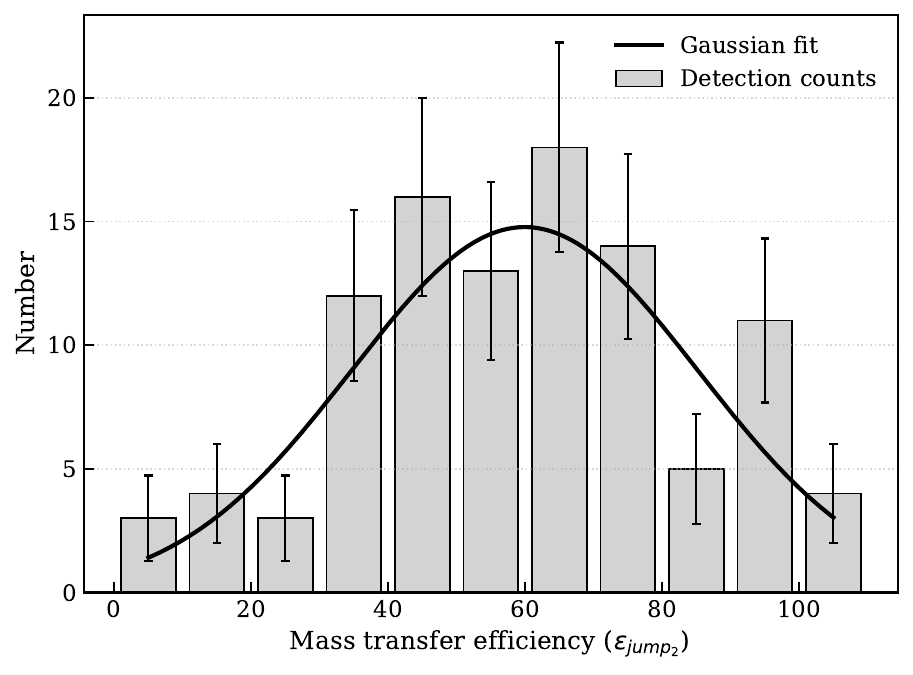}\hskip 0.2cm\includegraphics[width=0.235\textwidth]{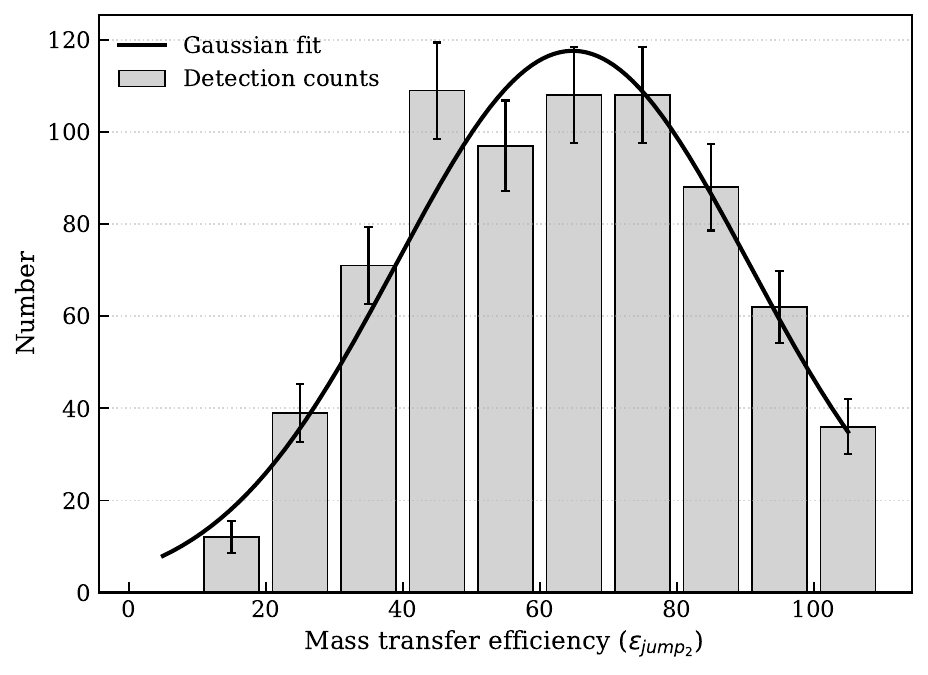}
    \vskip -0.1cm
    \caption{Distributions of mass transfer efficiencies in MHD (\textit{left panel}) and HD (\textit{right panel}) synthetic protoclusters. The error bars correspond to $\sqrt{N}$. Fitted by Gaussians, the distributions peak at $\overline{\epsilon_{\rm jump2}} \simeq 60\%$ for MHD and $\overline{\epsilon_{\rm jump2}} \simeq 65\%$ for HD protoclusters, respectively, and display a standard deviation of $25\%$.}
    \label{appendixfig:param-measured}
\end{figure}

\section{Core catalogs}
\cref{tab:MM1sources} lists the main characteristics of the cloud fragments extracted in W43-MM1, by \getsf and at all scales.  The complete catalog of sources identified in the five ALMA 3~mm images can be found at CDS.
\FloatBarrier
{\renewcommand{\arraystretch}{1}
\begin{table*}[ht]
\centering
\tiny
\begin{threeparttable}[c]
\caption{Catalog of sources identified by \getsf in the five ALMA 3~mm images of the W43-MM1 protocluster.}
\label{tab:MM1sources}
\begin{tabular}{cccccccccc}
\hline\hline
Node & \getsf & RA & Dec & $a_{\rm 3mm} \times b_{\rm 3mm}$ & PA &$S_{\rm 3mm}^{\rm peak}$ & $S_{\rm 3mm}^{\rm int}$ & Ascending family tree  \\
\multicolumn{2}{c}{number} & [J2000] & [J2000] &  [$\arcsec \times \arcsec$] & [°] & [mJy\,beam$^{-1}$] & [mJy] &  [node numbers] \\
\hline
\hline
\multicolumn{4}{l}{Image at 270~au resolution @99.00~GHz} & &  & & &\\
0 & 1\_1 & 18:47:46.84 & -01:54:29.33 & $0.08\times0.06$ & 135 & $ 2.96\pm0.03 $ & $ 10.35\pm0.08$ & [ 0 ; 57 ; 130 ; 177 ; 190 ] \\
1 & 1\_2 & 18:47:46.89 & -01:54:29.64 & $0.06\times0.05$ & 57 & $ 1.39\pm0.02 $ & $ 2.78\pm0.03$ & [ 1 ; 59 ; 140 ; 177 ; 190 ] \\
2 & 1\_3 & 18:47:46.25 & -01:54:33.31 & $0.06\times0.05$ & 81 & $ 0.84\pm0.02 $ & $ 1.31\pm0.02$ & [ 2 ; 67 ; 139 ; 180 ; - ] \\
3 & 1\_4 & 18:47:46.96 & -01:54:29.69 & $0.06\times0.05$ & 64 & $ 0.95\pm0.03 $ & $ 1.62\pm0.03$ & [ 3 ; 61 ; 155 ; - ; 190 ] \\
4 & 1\_5 & 18:47:46.97 & -01:54:26.49 & $0.06\times0.03$ & 70 & $ 1.67\pm0.06 $ & $ 1.84\pm0.05$ & [ 4 ; 70 ; (132 ; 143) ; 176 ; 189 ] \\
5 & 1\_6 & 18:47:46.37 & -01:54:33.54 & $0.06\times0.04$ & 71 & $ 1.1\pm0.04 $ & $ 1.42\pm0.04$ & [ 5 ; 75 ; 129 ; 178 ; - ] \\
6 & 1\_7 & 18:47:46.52 & -01:54:24.25 & $0.05\times0.04$ & 60 & $ 0.76\pm0.02 $ & $ 0.78\pm0.02$ & [ 6 ; 71 ; 142 ; 179 ; 191 ] \\
7 & 1\_8 & 18:47:44.77 & -01:54:38.08 & $0.05\times0.03$ & 70 & $ 0.7\pm0.02 $ & $ 0.67\pm0.02$ & [ 7 ; 69 ; 137 ; 186 ; - ] \\
8 & 1\_9 & 18:47:47.05 & -01:54:32.12 & $0.06\times0.04$ & 84 & $ 0.69\pm0.03 $ & $ 0.99\pm0.03$ & [ 8 ; 65 ; 145 ; - ; - ] \\
9 & 1\_10 & 18:47:46.47 & -01:54:32.60 & $0.06\times0.04$ & 82 & $ 0.72\pm0.03 $ & $ 0.91\pm0.03$ & [ 9 ; 84 ; 141 ; 181 ; - ] \\
10 & 1\_11 & 18:47:46.77 & -01:54:31.24 & $0.07\times0.05$ & 108 & $ 0.7\pm0.03 $ & $ 1.22\pm0.03$ & [ 10 ; 66 ; 135 ; 183 ; 190 ] \\
11 & 1\_12 & 18:47:46.53 & -01:54:23.21 & $0.06\times0.04$ & 90 & $ 0.75\pm0.03 $ & $ 1.19\pm0.03$ & [ 11 ; 64 ; 134 ; 179 ; 191 ] \\
12 & 1\_13 & 18:47:46.90 & -01:54:30.03 & $0.06\times0.04$ & 63 & $ 0.59\pm0.03 $ & $ 0.92\pm0.03$ & [ 12 ; 82 ; 140 ; 177 ; 190 ] \\
13 & 1\_14 & 18:47:47.03 & -01:54:31.06 & $0.06\times0.04$ & 57 & $ 0.56\pm0.03 $ & $ 0.8\pm0.03$ & [ 13 ; 68 ; 150 ; - ; - ] \\
14 & 1\_15 & 18:47:46.95 & -01:54:26.22 & $0.05\times0.04$ & 80 & $ 0.54\pm0.02 $ & $ 0.57\pm0.02$ & [ 14 ; 91 ; (132 ; 143) ; 176 ; 189 ] \\
15 & 1\_16 & 18:47:46.25 & -01:54:33.65 & $0.07\times0.04$ & 59 & $ 0.49\pm0.02 $ & $ 0.76\pm0.02$ & [ 15 ; 80 ; 139 ; 180 ; - ] \\
16 & 1\_17 & 18:47:47.01 & -01:54:30.77 & $0.06\times0.04$ & 68 & $ 0.55\pm0.03 $ & $ 0.61\pm0.03$ & [ 16 ; 85 ; 150 ; - ; 190 ] \\
17 & 1\_18 & 18:47:46.79 & -01:54:16.12 & $0.05\times0.04$ & 61 & $ 0.41\pm0.02 $ & $ 0.46\pm0.02$ & [ 17 ; 74 ; 138 ; 184 ; - ] \\
18 & 1\_19 & 18:47:46.56 & -01:54:32.13 & $0.06\times0.04$ & 32 & $ 0.43\pm0.02 $ & $ 0.55\pm0.02$ & [ 18 ; 81 ; 148 ; 181 ; - ] \\
19 & 1\_20 & 18:47:46.18 & -01:54:33.45 & $0.05\times0.05$ & 5 & $ 0.38\pm0.02 $ & $ 0.55\pm0.02$ & [ 19 ; 86 ; 133 ; 180 ; - ] \\
20 & 1\_21 & 18:47:47.26 & -01:54:29.74 & $0.07\times0.05$ & 71 & $ 0.9\pm0.07 $ & $ 1.61\pm0.08$ & [ 20 ; 58 ; 131 ; 182 ; - ] \\
21 & 1\_22 & 18:47:47.09 & -01:54:27.06 & $0.08\times0.06$ & 81 & $ 0.46\pm0.03 $ & $ 1.15\pm0.03$ & [ 21 ; 72 ; 146 ; 176 ; 189 ] \\
22 & 1\_23 & 18:47:46.36 & -01:54:33.46 & $0.08\times0.05$ & 62 & $ 0.91\pm0.13 $ & $ 1.86\pm0.14$ & [ 22 ; 62 ; 129 ; 178 ; - ] \\
23 & 1\_24 & 18:47:46.92 & -01:54:28.66 & $0.08\times0.05$ & 67 & $ 0.55\pm0.03 $ & $ 1.01\pm0.03$ & [ 23 ; 76 ; 158 ; 177 ; 190 ] \\
24 & 1\_25 & 18:47:47.02 & -01:54:26.78 & $0.07\times0.06$ & 127 & $ 0.58\pm0.06 $ & $ 1.21\pm0.06$ & [ 24 ; 87 ; 128 ; 176 ; 189 ] \\
25 & 1\_26 & 18:47:46.90 & -01:54:24.33 & $0.07\times0.06$ & 62 & $ 0.35\pm0.03 $ & $ 0.63\pm0.03$ & [ 25 ; 83 ; 160 ; - ; - ] \\
26 & 1\_27 & 18:47:46.16 & -01:54:33.30 & $0.05\times0.04$ & 64 & $ 0.5\pm0.05 $ & $ 0.6\pm0.04$ & [ 26 ; 73 ; 133 ; 180 ; - ] \\
27 & 1\_28 & 18:47:45.29 & -01:54:37.11 & $0.06\times0.04$ & 64 & $ 0.4\pm0.03 $ & $ 0.48\pm0.03$ & [ 27 ; 77 ; 136 ; 188 ; - ] \\
28 & 1\_29 & 18:47:46.93 & -01:54:27.18 & $0.14\times0.12$ & 62 & $ 0.19\pm0.02 $ & $ 1.38\pm0.05$ & [ 28 ; 89 ; 149 ; 176 ; (189 ; 190) ] \\
29 & 1\_30 & 18:47:47.03 & -01:54:26.90 & $0.07\times0.06$ & 73 & $ 0.59\pm0.08 $ & $ 1.16\pm0.06$ & [ 29 ; 63 ; 128 ; 176 ; 189 ] \\
30 & 1\_31 & 18:47:47.02 & -01:54:26.94 & $0.05\times0.03$ & 70 & $ 0.67\pm0.07 $ & $ 0.67\pm0.05$ & [ 30 ; 63 ; 128 ; 176 ; 189 ] \\
31 & 1\_32 & 18:47:46.58 & -01:54:32.05 & $0.06\times0.05$ & 61 & $ 0.28\pm0.03 $ & $ 0.48\pm0.03$ & [ 31 ; 88 ; 148 ; 181 ; - ] \\
32 & 1\_33 & 18:47:46.88 & -01:54:25.84 & $0.06\times0.05$ & 67 & $ 0.25\pm0.02 $ & $ 0.35\pm0.02$ & [ 32 ; 94 ; 157 ; - ; 189 ] \\
33 & 1\_34 & 18:47:47.02 & -01:54:27.00 & $0.08\times0.07$ & 102 & $ 0.46\pm0.06 $ & $ 1.22\pm0.07$ & [ 33 ; 63 ; 128 ; 176 ; 189 ] \\
34 & 1\_35 & 18:47:46.85 & -01:54:29.31 & $0.09\times0.08$ & 6 & $ 0.46\pm0.05 $ & $ 1.8\pm0.07$ & [ 34 ; - ; 130 ; 177 ; 190 ] \\
35 & 1\_36 & 18:47:46.98 & -01:54:26.57 & $0.07\times0.06$ & 23 & $ 0.26\pm0.04 $ & $ 0.62\pm0.04$ & [ 35 ; 70 ; (132 ; 143) ; 176 ; 189 ] \\
36 & 1\_37 & 18:47:46.15 & -01:54:33.35 & $0.07\times0.04$ & 67 & $ 0.37\pm0.05 $ & $ 0.49\pm0.05$ & [ 36 ; 78 ; 133 ; 180 ; - ] \\
37 & 1\_38 & 18:47:46.77 & -01:54:31.21 & $0.07\times0.05$ & 80 & $ 0.33\pm0.04 $ & $ 0.6\pm0.04$ & [ 37 ; 66 ; 135 ; 183 ; 190 ] \\
38 & 1\_39 & 18:47:46.53 & -01:54:23.14 & $0.08\times0.05$ & 67 & $ 0.34\pm0.03 $ & $ 0.55\pm0.03$ & [ 38 ; 64 ; 134 ; 179 ; 191 ] \\
39 & 1\_40 & 18:47:47.01 & -01:54:27.00 & $0.06\times0.05$ & 23 & $ 0.29\pm0.06 $ & $ 0.59\pm0.07$ & [ 39 ; 95 ; 128 ; 176 ; 189 ] \\
40 & 1\_41 & 18:47:46.85 & -01:54:29.55 & $0.09\times0.07$ & 160 & $ 0.18\pm0.03 $ & $ 0.51\pm0.03$ & [ 40 ; 101 ; 130 ; 177 ; 190 ] \\
41 & 1\_42 & 18:47:46.87 & -01:54:14.61 & $0.06\times0.03$ & 50 & $ 0.22\pm0.02 $ & $ 0.22\pm0.02$ & [ 41 ; 93 ; 151 ; - ; - ] \\
42 & 1\_43 & 18:47:45.22 & -01:54:38.81 & $0.06\times0.04$ & 87 & $ 0.17\pm0.02 $ & $ 0.18\pm0.02$ & [ 42 ; 98 ; 153 ; - ; - ] \\
43 & 1\_44 & 18:47:46.68 & -01:54:31.25 & $0.05\times0.03$ & 59 & $ 0.22\pm0.02 $ & $ 0.23\pm0.02$ & [ 43 ; 92 ; - ; 183 ; - ] \\
44 & 1\_45 & 18:47:47.27 & -01:54:29.67 & $0.06\times0.04$ & 83 & $ 0.36\pm0.07 $ & $ 0.51\pm0.07$ & [ 44 ; 79 ; 131 ; 182 ; - ] \\
45 & 1\_46 & 18:47:46.35 & -01:54:33.82 & $0.06\times0.04$ & 86 & $ 0.18\pm0.02 $ & $ 0.21\pm0.02$ & [ 45 ; 106 ; - ; 178 ; - ] \\
46 & 1\_47 & 18:47:47.03 & -01:54:26.85 & $0.05\times0.04$ & 80 & $ 0.3\pm0.06 $ & $ 0.34\pm0.05$ & [ 46 ; 63 ; 128 ; 176 ; 189 ] \\
47 & 1\_48 & 18:47:47.02 & -01:54:26.85 & $0.06\times0.05$ & 4 & $ 0.32\pm0.08 $ & $ 0.38\pm0.06$ & [ 47 ; 63 ; 128 ; 176 ; 189 ] \\
48 & 1\_49 & 18:47:45.05 & -01:54:42.13 & $0.05\times0.03$ & 72 & $ 0.21\pm0.03 $ & $ 0.18\pm0.02$ & [ 48 ; 96 ; 166 ; - ; - ] \\
49 & 1\_50 & 18:47:47.03 & -01:54:27.01 & $0.08\times0.07$ & 10 & $ 0.31\pm0.07 $ & $ 0.88\pm0.08$ & [ 49 ; 63 ; 128 ; 176 ; 189 ] \\
50 & 1\_51 & 18:47:46.76 & -01:54:16.80 & $0.05\times0.04$ & 70 & $ 0.13\pm0.02 $ & $ 0.17\pm0.03$ & [ 50 ; 99 ; - ; 184 ; - ] \\
51 & 1\_52 & 18:47:46.39 & -01:54:33.22 & $0.11\times0.09$ & 79 & $ 0.11\pm0.03 $ & $ 0.42\pm0.05$ & [ 51 ; 109 ; 129 ; 178 ; - ] \\
52 & 1\_53 & 18:47:46.93 & -01:54:27.23 & $0.07\times0.04$ & 64 & $ 0.18\pm0.02 $ & $ 0.23\pm0.02$ & [ 52 ; 89 ; 149 ; 176 ; (189 ; 190) ] \\
53 & 1\_54 & 18:47:46.72 & -01:54:17.55 & $0.1\times0.07$ & 4 & $ 0.12\pm0.02 $ & $ 0.23\pm0.03$ & [ 53 ; 111 ; 172 ; - ; - ] \\
54 & 1\_55 & 18:47:46.54 & -01:54:23.06 & $0.1\times0.05$ & 62 & $ 0.18\pm0.03 $ & $ 0.33\pm0.03$ & [ 54 ; - ; 134 ; 179 ; 191 ] \\
55 & 1\_56 & 18:47:47.09 & -01:54:27.01 & $0.06\times0.04$ & 85 & $ 0.2\pm0.03 $ & $ 0.17\pm0.03$ & [ 55 ; 72 ; 146 ; 176 ; 189 ] \\
56 & 1\_57 & 18:47:46.56 & -01:54:20.95 & $0.12\times0.08$ & 59 & $ 0.09\pm0.03 $ & $ 0.13\pm0.04$ & [ 56 ; 108 ; 156 ; - ; 191 ] \\
&      \\ 
\hline
\end{tabular}
\begin{tablenotes}[para,flushleft]
Notes: Number given to the sources in the hierarchical network (nodes) and in the \getsf catalogs of the five images. RA, right ascension; Dec, declination; $a$ and $b$, major and minor sizes at half maximum; PA, counterclockwise ellipse orientation from north to east; $S^{\rm peak}$ and $S^{\rm int}$, peak and integrated fluxes at the given central frequency and for beam sizes of Col.~5; 
Ascending nodes of the source family tree. The complete catalog of sources identified in the five ALMA 3~mm images of the W43-MM1 protocluster can be found at CDS.
\end{tablenotes}
\end{threeparttable}
\end{table*}
}

\setcounter{table}{0}
{\renewcommand{\arraystretch}{1}
\begin{table*}[ht]
\centering
\tiny
\begin{threeparttable}[c]
\caption{(Continued).}
\begin{tabular}{cccccccccc}
\hline\hline
Node & \getsf & RA & Dec & $a_{\rm 3mm} \times b_{\rm 3mm}$ & PA &$S_{\rm 3mm}^{\rm peak}$ & $S_{\rm 3mm}^{\rm int}$ & Ascending family tree  \\
\multicolumn{2}{c}{number} & [J2000] & [J2000] &  [$\arcsec \times \arcsec$] & [°] & [mJy\,beam$^{-1}$] & [mJy] &  [node numbers] \\
\hline
\hline
\multicolumn{4}{l}{Image at 650~au resolution @85.10~GHz} & & & & \\
57 & 2\_1 & 18:47:46.84 & -01:54:29.33 & $0.15\times0.12$ & 92 & $ 5.04\pm0.07 $ & $ 24.51\pm0.14$ & [ - ; 57 ; 130 ; 177 ; 190 ] \\
58 & 2\_2 & 18:47:47.25 & -01:54:29.74 & $0.14\times0.12$ & 85 & $ 1.69\pm0.02 $ & $ 5.82\pm0.02$ & [ - ; 58 ; 131 ; 182 ; - ] \\
59 & 2\_3 & 18:47:46.89 & -01:54:29.64 & $0.13\times0.11$ & 85 & $ 1.93\pm0.04 $ & $ 5.76\pm0.04$ & [ - ; 59 ; 140 ; 177 ; 190 ] \\
60 & 2\_4 & 18:47:46.37 & -01:54:33.32 & $0.15\times0.12$ & 72 & $ 2.09\pm0.04 $ & $ 7.78\pm0.05$ & [ - ; 60 ; 129 ; 178 ; - ] \\
61 & 2\_5 & 18:47:46.96 & -01:54:29.69 & $0.12\times0.11$ & 79 & $ 1.19\pm0.02 $ & $ 3.26\pm0.02$ & [ - ; 61 ; 155 ; - ; 190 ] \\
62 & 2\_6 & 18:47:46.36 & -01:54:33.47 & $0.13\times0.12$ & 74 & $ 1.75\pm0.04 $ & $ 5.16\pm0.04$ & [ - ; 62 ; 129 ; 178 ; - ] \\
63 & 2\_7 & 18:47:47.03 & -01:54:26.95 & $0.21\times0.2$ & 97 & $ 2.08\pm0.1 $ & $ 18.4\pm0.2$ & [ - ; 63 ; 128 ; 176 ; 189 ] \\
64 & 2\_8 & 18:47:46.53 & -01:54:23.18 & $0.14\times0.11$ & 49 & $ 0.94\pm0.02 $ & $ 2.98\pm0.02$ & [ - ; 64 ; 134 ; 179 ; 191 ] \\
65 & 2\_9 & 18:47:47.05 & -01:54:32.12 & $0.13\times0.11$ & 1 & $ 0.87\pm0.01 $ & $ 2.79\pm0.02$ & [ - ; 65 ; 145 ; - ; - ] \\
66 & 2\_10 & 18:47:46.77 & -01:54:31.23 & $0.16\times0.11$ & 99 & $ 1.39\pm0.04 $ & $ 5.21\pm0.06$ & [ - ; 66 ; 135 ; 183 ; 190 ] \\
67 & 2\_11 & 18:47:46.25 & -01:54:33.31 & $0.12\times0.11$ & 105 & $ 0.95\pm0.02 $ & $ 2.46\pm0.02$ & [ - ; 67 ; 139 ; 180 ; - ] \\
68 & 2\_12 & 18:47:47.03 & -01:54:31.05 & $0.12\times0.12$ & 91 & $ 0.82\pm0.02 $ & $ 2.35\pm0.02$ & [ - ; 68 ; 150 ; - ; - ] \\
69 & 2\_13 & 18:47:44.78 & -01:54:38.08 & $0.11\times0.1$ & 81 & $ 0.82\pm0.02 $ & $ 1.92\pm0.02$ & [ - ; 69 ; 137 ; 186 ; - ] \\
70 & 2\_14 & 18:47:46.97 & -01:54:26.50 & $0.13\times0.12$ & 20 & $ 1.97\pm0.1 $ & $ 7.14\pm0.13$ & [ - ; 70 ; (132 ; 143) ; 176 ; 189 ] \\
71 & 2\_15 & 18:47:46.52 & -01:54:24.25 & $0.12\times0.1$ & 101 & $ 0.73\pm0.02 $ & $ 1.83\pm0.02$ & [ - ; 71 ; 142 ; 179 ; 191 ] \\
72 & 2\_16 & 18:47:47.09 & -01:54:27.06 & $0.15\times0.11$ & 121 & $ 0.79\pm0.03 $ & $ 2.92\pm0.03$ & [ - ; 72 ; 146 ; 176 ; 189 ] \\
73 & 2\_17 & 18:47:46.16 & -01:54:33.31 & $0.13\times0.12$ & 120 & $ 0.7\pm0.02 $ & $ 2.12\pm0.02$ & [ - ; 73 ; 133 ; 180 ; - ] \\
74 & 2\_18 & 18:47:46.79 & -01:54:16.09 & $0.14\times0.13$ & 117 & $ 0.64\pm0.02 $ & $ 3.07\pm0.02$ & [ - ; 74 ; 138 ; 184 ; - ] \\
75 & 2\_19 & 18:47:46.37 & -01:54:33.54 & $0.14\times0.1$ & 88 & $ 1.07\pm0.03 $ & $ 3.14\pm0.04$ & [ - ; 75 ; 129 ; 178 ; - ] \\
76 & 2\_20 & 18:47:46.92 & -01:54:28.66 & $0.13\times0.1$ & 85 & $ 0.84\pm0.03 $ & $ 2.38\pm0.03$ & [ - ; 76 ; 158 ; 177 ; 190 ] \\
77 & 2\_21 & 18:47:45.29 & -01:54:37.11 & $0.12\times0.11$ & 24 & $ 0.54\pm0.01 $ & $ 1.66\pm0.02$ & [ - ; 77 ; 136 ; 188 ; - ] \\
78 & 2\_22 & 18:47:46.15 & -01:54:33.33 & $0.17\times0.13$ & 44 & $ 0.76\pm0.02 $ & $ 3.94\pm0.03$ & [ - ; 78 ; 133 ; 180 ; - ] \\
79 & 2\_23 & 18:47:47.27 & -01:54:29.66 & $0.16\times0.11$ & 96 & $ 0.76\pm0.02 $ & $ 2.55\pm0.02$ & [ - ; 79 ; 131 ; 182 ; - ] \\
80 & 2\_24 & 18:47:46.25 & -01:54:33.65 & $0.12\times0.11$ & 88 & $ 0.57\pm0.02 $ & $ 1.44\pm0.02$ & [ - ; 80 ; 139 ; 180 ; - ] \\
81 & 2\_25 & 18:47:46.56 & -01:54:32.12 & $0.13\times0.12$ & 73 & $ 0.6\pm0.02 $ & $ 2.15\pm0.02$ & [ - ; 81 ; 148 ; 181 ; - ] \\
82 & 2\_26 & 18:47:46.90 & -01:54:30.04 & $0.13\times0.11$ & 86 & $ 0.75\pm0.03 $ & $ 2.45\pm0.04$ & [ - ; 82 ; 140 ; 177 ; 190 ] \\
83 & 2\_27 & 18:47:46.90 & -01:54:24.33 & $0.13\times0.11$ & 91 & $ 0.54\pm0.02 $ & $ 1.73\pm0.03$ & [ - ; 83 ; 160 ; - ; - ] \\
84 & 2\_28 & 18:47:46.47 & -01:54:32.60 & $0.14\times0.1$ & 99 & $ 0.78\pm0.04 $ & $ 2.33\pm0.04$ & [ - ; 84 ; 141 ; 181 ; - ] \\
85 & 2\_29 & 18:47:47.01 & -01:54:30.76 & $0.13\times0.11$ & 170 & $ 0.59\pm0.03 $ & $ 1.8\pm0.03$ & [ - ; 85 ; 150 ; - ; 190 ] \\
86 & 2\_30 & 18:47:46.18 & -01:54:33.45 & $0.12\times0.1$ & 97 & $ 0.42\pm0.02 $ & $ 0.97\pm0.01$ & [ - ; 86 ; 133 ; 180 ; - ] \\
87 & 2\_31 & 18:47:47.02 & -01:54:26.77 & $0.14\times0.13$ & 56 & $ 1.12\pm0.09 $ & $ 4.22\pm0.1$ & [ - ; 87 ; 128 ; 176 ; 189 ] \\
88 & 2\_32 & 18:47:46.58 & -01:54:32.05 & $0.15\times0.11$ & 54 & $ 0.47\pm0.02 $ & $ 1.82\pm0.03$ & [ - ; 88 ; 148 ; 181 ; - ] \\
89 & 2\_33 & 18:47:46.93 & -01:54:27.18 & $0.19\times0.13$ & 73 & $ 0.58\pm0.04 $ & $ 3.14\pm0.08$ & [ - ; 89 ; 149 ; 176 ; (189 ; 190) ] \\
90 & 2\_34 & 18:47:46.54 & -01:54:32.50 & $0.21\times0.14$ & 122 & $ 0.24\pm0.01 $ & $ 1.32\pm0.02$ & [ - ; 90 ; - ; 181 ; - ] \\
91 & 2\_35 & 18:47:46.95 & -01:54:26.22 & $0.11\times0.1$ & 113 & $ 0.47\pm0.05 $ & $ 1.03\pm0.04$ & [ - ; 91 ; (132 ; 143) ; 176 ; 189 ] \\
92 & 2\_36 & 18:47:46.68 & -01:54:31.26 & $0.12\times0.11$ & 107 & $ 0.27\pm0.02 $ & $ 0.71\pm0.01$ & [ - ; 92 ; - ; 183 ; - ] \\
93 & 2\_37 & 18:47:46.87 & -01:54:14.60 & $0.12\times0.11$ & 68 & $ 0.24\pm0.01 $ & $ 0.57\pm0.01$ & [ - ; 93 ; 151 ; - ; - ] \\
94 & 2\_38 & 18:47:46.88 & -01:54:25.82 & $0.16\times0.11$ & 150 & $ 0.33\pm0.02 $ & $ 1.28\pm0.03$ & [ - ; 94 ; 157 ; - ; 189 ] \\
95 & 2\_39 & 18:47:47.01 & -01:54:27.00 & $0.19\times0.15$ & 124 & $ 0.81\pm0.1 $ & $ 4.76\pm0.13$ & [ - ; 95 ; 128 ; 176 ; 189 ] \\
96 & 2\_40 & 18:47:45.05 & -01:54:42.14 & $0.1\times0.1$ & 155 & $ 0.2\pm0.02 $ & $ 0.38\pm0.01$ & [ - ; 96 ; 166 ; - ; - ] \\
97 & 2\_41 & 18:47:47.03 & -01:54:24.95 & $0.19\times0.16$ & 177 & $ 0.15\pm0.01 $ & $ 0.81\pm0.02$ & [ - ; 97 ; - ; - ; 189 ] \\
98 & 2\_42 & 18:47:45.22 & -01:54:38.82 & $0.14\times0.1$ & 147 & $ 0.2\pm0.01 $ & $ 0.58\pm0.01$ & [ - ; 98 ; 153 ; - ; - ] \\
99 & 2\_43 & 18:47:46.76 & -01:54:16.81 & $0.13\times0.1$ & 99 & $ 0.14\pm0.01 $ & $ 0.36\pm0.01$ & [ - ; 99 ; - ; 184 ; - ] \\
100 & 2\_44 & 18:47:47.32 & -01:54:12.86 & $0.11\times0.1$ & 115 & $ 0.17\pm0.01 $ & $ 0.33\pm0.01$ & [ - ; 100 ; 164 ; - ; - ] \\
101 & 2\_45 & 18:47:46.85 & -01:54:29.58 & $0.21\times0.15$ & 167 & $ 0.47\pm0.05 $ & $ 2.33\pm0.05$ & [ - ; 101 ; 130 ; 177 ; 190 ] \\
102 & 2\_46 & 18:47:46.83 & -01:54:26.85 & $0.14\times0.13$ & 177 & $ 0.13\pm0.01 $ & $ 0.46\pm0.01$ & [ - ; 102 ; - ; - ; 190 ] \\
103 & 2\_47 & 18:47:46.89 & -01:54:30.60 & $0.17\times0.11$ & 146 & $ 0.24\pm0.02 $ & $ 0.94\pm0.02$ & [ - ; 103 ; - ; - ; 190 ] \\
104 & 2\_48 & 18:47:45.25 & -01:54:39.80 & $0.12\times0.11$ & 102 & $ 0.16\pm0.01 $ & $ 0.45\pm0.01$ & [ - ; 104 ; 162 ; - ; - ] \\
105 & 2\_49 & 18:47:46.97 & -01:54:32.07 & $0.13\times0.1$ & 157 & $ 0.2\pm0.02 $ & $ 0.51\pm0.02$ & [ - ; 105 ; 167 ; - ; - ] \\
106 & 2\_50 & 18:47:46.35 & -01:54:33.83 & $0.16\times0.12$ & 32 & $ 0.21\pm0.02 $ & $ 0.87\pm0.03$ & [ - ; 106 ; - ; 178 ; - ] \\
107 & 2\_51 & 18:47:47.35 & -01:54:13.42 & $0.11\times0.1$ & 60 & $ 0.15\pm0.01 $ & $ 0.33\pm0.01$ & [ - ; 107 ; 163 ; - ; - ] \\
108 & 2\_52 & 18:47:46.57 & -01:54:20.96 & $0.16\times0.1$ & 55 & $ 0.2\pm0.01 $ & $ 0.7\pm0.02$ & [ - ; 108 ; 156 ; - ; 191 ] \\
109 & 2\_53 & 18:47:46.38 & -01:54:33.19 & $0.24\times0.19$ & 79 & $ 0.22\pm0.04 $ & $ 1.7\pm0.05$ & [ - ; 109 ; 129 ; 178 ; - ] \\
110 & 2\_54 & 18:47:47.01 & -01:54:30.32 & $0.14\times0.12$ & 145 & $ 0.17\pm0.03 $ & $ 0.56\pm0.02$ & [ - ; 110 ; - ; - ; 190 ] \\
111 & 2\_55 & 18:47:46.72 & -01:54:17.55 & $0.12\times0.12$ & 85 & $ 0.11\pm0.01 $ & $ 0.29\pm0.01$ & [ - ; 111 ; 172 ; - ; - ] \\
112 & 2\_56 & 18:47:45.27 & -01:54:40.21 & $0.11\times0.09$ & 131 & $ 0.15\pm0.02 $ & $ 0.3\pm0.01$ & [ - ; 112 ; 162 ; - ; - ] \\
113 & 2\_57 & 18:47:46.59 & -01:54:20.54 & $0.11\times0.11$ & 53 & $ 0.09\pm0.01 $ & $ 0.21\pm0.01$ & [ - ; 113 ; - ; - ; - ] \\
114 & 2\_58 & 18:47:46.64 & -01:54:19.42 & $0.11\times0.09$ & 150 & $ 0.1\pm0.01 $ & $ 0.2\pm0.01$ & [ - ; 114 ; 169 ; - ; - ] \\
115 & 2\_59 & 18:47:46.50 & -01:54:28.75 & $0.14\times0.09$ & 130 & $ 0.13\pm0.02 $ & $ 0.33\pm0.02$ & [ - ; 115 ; 161 ; - ; - ] \\
116 & 2\_60 & 18:47:46.80 & -01:54:21.62 & $0.13\times0.11$ & 49 & $ 0.1\pm0.01 $ & $ 0.23\pm0.01$ & [ - ; 116 ; 159 ; - ; - ] \\
117 & 2\_61 & 18:47:47.18 & -01:54:21.58 & $0.28\times0.22$ & 89 & $ 0.07\pm0.01 $ & $ 0.68\pm0.01$ & [ - ; 117 ; 175 ; - ; - ] \\
118 & 2\_62 & 18:47:46.35 & -01:54:29.64 & $0.21\times0.17$ & 65 & $ 0.07\pm0.01 $ & $ 0.43\pm0.02$ & [ - ; 118 ; 168 ; - ; - ] \\
119 & 2\_63 & 18:47:46.98 & -01:54:26.11 & $0.14\times0.12$ & 147 & $ 0.19\pm0.08 $ & $ 0.54\pm0.06$ & [ - ; 119 ; (132 ; 143) ; 176 ; 189 ] \\
120 & 2\_64 & 18:47:45.41 & -01:54:37.53 & $0.11\times0.09$ & 94 & $ 0.09\pm0.01 $ & $ 0.19\pm0.01$ & [ - ; 120 ; 165 ; - ; - ] \\
121 & 2\_65 & 18:47:47.00 & -01:54:25.69 & $0.17\times0.16$ & 127 & $ 0.13\pm0.07 $ & $ 0.56\pm0.06$ & [ - ; 121 ; 143 ; 176 ; 189 ] \\
122 & 2\_66 & 18:47:47.06 & -01:54:26.87 & $0.12\times0.11$ & 40 & $ 0.08\pm0.03 $ & $ 0.2\pm0.02$ & [ - ; 122 ; - ; 176 ; 189 ] \\
123 & 2\_67 & 18:47:46.94 & -01:54:26.83 & $0.18\times0.15$ & 11 & $ 0.09\pm0.03 $ & $ 0.38\pm0.03$ & [ - ; 123 ; - ; 176 ; 189 ] \\
124 & 2\_68 & 18:47:46.49 & -01:54:32.46 & $0.13\times0.11$ & 45 & $ 0.08\pm0.03 $ & $ 0.2\pm0.03$ & [ - ; 124 ; 141 ; 181 ; - ] \\
125 & 2\_69 & 18:47:46.74 & -01:54:31.49 & $0.21\times0.19$ & 33 & $ 0.09\pm0.03 $ & $ 0.59\pm0.04$ & [ - ; 125 ; - ; 183 ; - ] \\
126 & 2\_70 & 18:47:46.96 & -01:54:23.84 & $0.28\times0.18$ & 46 & $ 0.05\pm0.02 $ & $ 0.3\pm0.02$ & [ - ; 126 ; - ; - ; - ] \\
127 & 2\_71 & 18:47:46.74 & -01:54:31.29 & $0.14\times0.11$ & 86 & $ 0.1\pm0.04 $ & $ 0.3\pm0.03$ & [ - ; 127 ; 135 ; 183 ; 190 ] \\
 &      \\ 
\hline
\end{tabular}
\end{threeparttable}
\end{table*}
}

\setcounter{table}{0}
{\renewcommand{\arraystretch}{1}
\begin{table*}[ht]
\centering
\tiny
\begin{threeparttable}[c]
\caption{(Continued).}
\begin{tabular}{cccccccccc}
\hline\hline
Node & \getsf & RA & Dec & $a_{\rm 3mm} \times b_{\rm 3mm}$ & PA &$S_{\rm 3mm}^{\rm peak}$ & $S_{\rm 3mm}^{\rm int}$ & Ascending family tree  \\
\multicolumn{2}{c}{number} & [J2000] & [J2000] &  [$\arcsec \times \arcsec$] & [°] & [mJy\,beam$^{-1}$] & [mJy] &  [node numbers] \\
\hline
\hline
\multicolumn{4}{l}{Image at 2.4~kau resolution @100.69~GHz} & & & & \\
128 & 3\_1 & 18:47:47.02 & -01:54:26.91 & $0.61\times0.46$ & 104 & $ 16.6\pm0.28 $ & $ 37.38\pm0.33$ & [ - ; - ; 128 ; 176 ; 189 ] \\
129 & 3\_2 & 18:47:46.36 & -01:54:33.39 & $0.53\times0.47$ & 79 & $ 8.21\pm0.13 $ & $ 18.06\pm0.17$ & [ - ; - ; 129 ; 178 ; - ] \\
130 & 3\_3 & 18:47:46.84 & -01:54:29.30 & $0.5\times0.36$ & 113 & $ 14.06\pm0.4 $ & $ 21.14\pm0.41$ & [ - ; - ; 130 ; 177 ; 190 ] \\
131 & 3\_4 & 18:47:47.26 & -01:54:29.70 & $0.53\times0.32$ & 102 & $ 4.43\pm0.04 $ & $ 6.02\pm0.04$ & [ - ; - ; 131 ; 182 ; - ] \\
132 & 3\_5 & 18:47:46.97 & -01:54:26.49 & $0.6\times0.56$ & 85 & $ 7.3\pm0.26 $ & $ 18.67\pm0.3$ & [ - ; - ; 132 ; 176 ; 189 ] \\
133 & 3\_6 & 18:47:46.15 & -01:54:33.33 & $0.59\times0.35$ & 106 & $ 3.42\pm0.08 $ & $ 5.42\pm0.08$ & [ - ; - ; 133 ; 180 ; - ] \\
134 & 3\_7 & 18:47:46.53 & -01:54:23.13 & $0.55\times0.33$ & 107 & $ 3.4\pm0.08 $ & $ 5.63\pm0.1$ & [ - ; - ; 134 ; 179 ; 191 ] \\
135 & 3\_8 & 18:47:46.77 & -01:54:31.21 & $0.54\times0.36$ & 92 & $ 3.55\pm0.17 $ & $ 5.59\pm0.18$ & [ - ; - ; 135 ; 183 ; 190 ] \\
136 & 3\_9 & 18:47:45.29 & -01:54:37.08 & $0.49\times0.34$ & 99 & $ 1.25\pm0.03 $ & $ 1.64\pm0.03$ & [ - ; - ; 136 ; 188 ; - ] \\
137 & 3\_10 & 18:47:44.77 & -01:54:38.03 & $0.48\times0.32$ & 108 & $ 1.11\pm0.03 $ & $ 1.44\pm0.03$ & [ - ; - ; 137 ; 186 ; - ] \\
138 & 3\_11 & 18:47:46.79 & -01:54:16.07 & $0.52\times0.33$ & 103 & $ 1.5\pm0.09 $ & $ 1.96\pm0.08$ & [ - ; - ; 138 ; 184 ; - ] \\
139 & 3\_12 & 18:47:46.25 & -01:54:33.38 & $0.55\times0.38$ & 128 & $ 1.65\pm0.11 $ & $ 2.57\pm0.1$ & [ - ; - ; 139 ; 180 ; - ] \\
140 & 3\_13 & 18:47:46.89 & -01:54:29.66 & $0.58\times0.41$ & 128 & $ 3.4\pm0.35 $ & $ 5.78\pm0.35$ & [ - ; - ; 140 ; 177 ; 190 ] \\
141 & 3\_14 & 18:47:46.48 & -01:54:32.57 & $0.51\times0.37$ & 99 & $ 2.27\pm0.25 $ & $ 3.33\pm0.22$ & [ - ; - ; 141 ; 181 ; - ] \\
142 & 3\_15 & 18:47:46.52 & -01:54:24.24 & $0.5\times0.33$ & 105 & $ 1.35\pm0.09 $ & $ 1.7\pm0.08$ & [ - ; - ; 142 ; 179 ; 191 ] \\
143 & 3\_16 & 18:47:46.99 & -01:54:25.91 & $0.88\times0.81$ & 130 & $ 3.05\pm0.25 $ & $ 14.51\pm0.33$ & [ - ; - ; 143 ; 176 ; 189 ] \\
144 & 3\_17 & 18:47:44.77 & -01:54:45.27 & $0.57\times0.4$ & 98 & $ 0.71\pm0.06 $ & $ 1.33\pm0.07$ & [ - ; - ; 144 ; 187 ; - ] \\
145 & 3\_18 & 18:47:47.05 & -01:54:32.15 & $0.48\times0.3$ & 104 & $ 1.48\pm0.13 $ & $ 1.64\pm0.1$ & [ - ; - ; 145 ; - ; - ] \\
146 & 3\_19 & 18:47:47.10 & -01:54:27.06 & $0.56\times0.4$ & 100 & $ 1.8\pm0.23 $ & $ 2.48\pm0.18$ & [ - ; - ; 146 ; 176 ; 189 ] \\
147 & 3\_20 & 18:47:44.94 & -01:54:42.89 & $0.58\times0.37$ & 105 & $ 0.45\pm0.03 $ & $ 0.71\pm0.03$ & [ - ; - ; 147 ; - ; - ] \\
148 & 3\_21 & 18:47:46.57 & -01:54:32.06 & $0.56\times0.35$ & 89 & $ 1.96\pm0.26 $ & $ 2.75\pm0.24$ & [ - ; - ; 148 ; 181 ; - ] \\
149 & 3\_22 & 18:47:46.93 & -01:54:27.15 & $0.47\times0.33$ & 103 & $ 1.8\pm0.29 $ & $ 2.08\pm0.22$ & [ - ; - ; 149 ; 176 ; (189 ; 190) ] \\
150 & 3\_23 & 18:47:47.02 & -01:54:30.86 & $0.63\times0.44$ & 130 & $ 1.71\pm0.36 $ & $ 3.44\pm0.34$ & [ - ; - ; 150 ; - ; - ] \\
151 & 3\_24 & 18:47:46.87 & -01:54:14.59 & $0.51\times0.31$ & 92 & $ 0.61\pm0.05 $ & $ 0.75\pm0.05$ & [ - ; - ; 151 ; - ; - ] \\
152 & 3\_26 & 18:47:46.96 & -01:54:13.00 & $0.76\times0.55$ & 89 & $ 0.39\pm0.05 $ & $ 1.15\pm0.07$ & [ - ; - ; 152 ; 185 ; - ] \\
153 & 3\_27 & 18:47:45.22 & -01:54:38.81 & $0.48\times0.33$ & 116 & $ 0.48\pm0.06 $ & $ 0.61\pm0.05$ & [ - ; - ; 153 ; - ; - ] \\
154 & 3\_28 & 18:47:47.01 & -01:54:16.80 & $0.56\times0.33$ & 107 & $ 0.45\pm0.04 $ & $ 0.64\pm0.04$ & [ - ; - ; 154 ; - ; - ] \\
155 & 3\_29 & 18:47:46.96 & -01:54:29.67 & $0.5\times0.28$ & 114 & $ 1.95\pm0.41 $ & $ 2.15\pm0.32$ & [ - ; - ; 155 ; - ; 190 ] \\
156 & 3\_30 & 18:47:46.57 & -01:54:20.96 & $0.48\times0.42$ & 76 & $ 0.49\pm0.1 $ & $ 0.76\pm0.09$ & [ - ; - ; 156 ; - ; 191 ] \\
157 & 3\_31 & 18:47:46.88 & -01:54:25.76 & $0.48\times0.32$ & 117 & $ 0.97\pm0.24 $ & $ 1.11\pm0.19$ & [ - ; - ; 157 ; - ; 189 ] \\
158 & 3\_32 & 18:47:46.92 & -01:54:28.63 & $0.5\times0.3$ & 107 & $ 1.69\pm0.34 $ & $ 1.88\pm0.27$ & [ - ; - ; 158 ; 177 ; 190 ] \\
159 & 3\_33 & 18:47:46.80 & -01:54:21.58 & $0.42\times0.32$ & 120 & $ 0.41\pm0.04 $ & $ 0.46\pm0.03$ & [ - ; - ; 159 ; - ; - ] \\
160 & 3\_34 & 18:47:46.90 & -01:54:24.28 & $0.53\times0.31$ & 108 & $ 1.05\pm0.2 $ & $ 1.24\pm0.16$ & [ - ; - ; 160 ; - ; - ] \\
161 & 3\_35 & 18:47:46.51 & -01:54:28.73 & $0.54\times0.33$ & 103 & $ 0.43\pm0.06 $ & $ 0.59\pm0.05$ & [ - ; - ; 161 ; - ; - ] \\
162 & 3\_36 & 18:47:45.26 & -01:54:39.97 & $0.83\times0.53$ & 116 & $ 0.24\pm0.06 $ & $ 0.71\pm0.06$ & [ - ; - ; 162 ; - ; - ] \\
163 & 3\_37 & 18:47:47.35 & -01:54:13.42 & $0.46\times0.37$ & 119 & $ 0.3\pm0.04 $ & $ 0.38\pm0.04$ & [ - ; - ; 163 ; - ; - ] \\
164 & 3\_38 & 18:47:47.33 & -01:54:12.85 & $0.43\times0.32$ & 92 & $ 0.31\pm0.04 $ & $ 0.34\pm0.04$ & [ - ; - ; 164 ; - ; - ] \\
165 & 3\_39 & 18:47:45.41 & -01:54:37.49 & $0.47\times0.31$ & 111 & $ 0.28\pm0.03 $ & $ 0.32\pm0.03$ & [ - ; - ; 165 ; - ; - ] \\
166 & 3\_40 & 18:47:45.05 & -01:54:42.12 & $0.46\times0.29$ & 120 & $ 0.31\pm0.04 $ & $ 0.32\pm0.03$ & [ - ; - ; 166 ; - ; - ] \\
167 & 3\_41 & 18:47:46.97 & -01:54:32.12 & $0.46\times0.32$ & 113 & $ 0.77\pm0.19 $ & $ 0.85\pm0.15$ & [ - ; - ; 167 ; - ; - ] \\
168 & 3\_42 & 18:47:46.34 & -01:54:29.60 & $0.51\times0.34$ & 123 & $ 0.27\pm0.05 $ & $ 0.37\pm0.04$ & [ - ; - ; 168 ; - ; - ] \\
169 & 3\_43 & 18:47:46.64 & -01:54:19.45 & $0.55\times0.37$ & 103 & $ 0.31\pm0.06 $ & $ 0.41\pm0.06$ & [ - ; - ; 169 ; - ; - ] \\
170 & 3\_44 & 18:47:44.77 & -01:54:44.18 & $0.47\times0.37$ & 118 & $ 0.27\pm0.06 $ & $ 0.4\pm0.05$ & [ - ; - ; 170 ; 187 ; - ] \\
171 & 3\_46 & 18:47:44.61 & -01:54:42.14 & $0.56\times0.45$ & 117 & $ 0.16\pm0.04 $ & $ 0.31\pm0.04$ & [ - ; - ; 171 ; - ; - ] \\
172 & 3\_48 & 18:47:46.72 & -01:54:17.49 & $0.43\times0.34$ & 77 & $ 0.37\pm0.09 $ & $ 0.41\pm0.07$ & [ - ; - ; 172 ; - ; - ] \\
173 & 3\_49 & 18:47:44.83 & -01:54:42.92 & $0.49\times0.36$ & 93 & $ 0.19\pm0.04 $ & $ 0.23\pm0.04$ & [ - ; - ; 173 ; - ; - ] \\
174 & 3\_51 & 18:47:46.36 & -01:54:16.83 & $0.46\times0.35$ & 7 & $ 0.18\pm0.05 $ & $ 0.18\pm0.05$ & [ - ; - ; 174 ; - ; - ] \\
175 & 3\_52 & 18:47:47.18 & -01:54:21.56 & $0.54\times0.27$ & 100 & $ 0.25\pm0.07 $ & $ 0.26\pm0.06$ & [ - ; - ; 175 ; - ; - ] \\

\hline
\multicolumn{4}{l}{Image at 8~kau resolution @99.00~GHz} & & & & \\
176 & 4\_1 & 18:47:47.01 & -01:54:26.73 & $1.82\times1.35$ & 100 & $ 53.58\pm2.52 $ & $ 83.34\pm2.57$ & [ - ; - ; - ; 176 ; 189 ] \\
177 & 4\_2 & 18:47:46.86 & -01:54:29.31 & $1.59\times1.21$ & 97 & $ 30.39\pm2.94 $ & $ 35.75\pm2.28$ & [ - ; - ; - ; 177 ; 190 ] \\
178 & 4\_3 & 18:47:46.37 & -01:54:33.37 & $1.75\times1.15$ & 80 & $ 21.35\pm0.78 $ & $ 26.57\pm0.62$ & [ - ; - ; - ; 178 ; - ] \\
179 & 4\_4 & 18:47:46.54 & -01:54:23.15 & $1.69\times1.51$ & 66 & $ 8.68\pm0.76 $ & $ 14.81\pm0.81$ & [ - ; - ; - ; 179 ; 191 ] \\
180 & 4\_5 & 18:47:46.17 & -01:54:33.34 & $2.1\times1.09$ & 88 & $ 8.82\pm0.57 $ & $ 11.56\pm0.89$ & [ - ; - ; - ; 180 ; - ] \\
181 & 4\_6 & 18:47:46.55 & -01:54:32.19 & $2.31\times1.39$ & 77 & $ 10.33\pm1.0 $ & $ 16.82\pm0.78$ & [ - ; - ; - ; 181 ; - ] \\
182 & 4\_7 & 18:47:47.28 & -01:54:29.68 & $1.58\times0.96$ & 90 & $ 6.25\pm1.48 $ & $ 6.2\pm1.15$ & [ - ; - ; - ; 182 ; - ] \\
183 & 4\_8 & 18:47:46.75 & -01:54:31.39 & $2.02\times1.21$ & 65 & $ 6.9\pm2.22 $ & $ 8.43\pm1.72$ & [ - ; - ; - ; 183 ; - ] \\
184 & 4\_9 & 18:47:46.80 & -01:54:16.10 & $1.37\times1.11$ & 54 & $ 2.76\pm0.46 $ & $ 3.34\pm0.36$ & [ - ; - ; - ; 184 ; - ] \\
185 & 4\_10 & 18:47:46.97 & -01:54:12.99 & $1.47\times1.03$ & 86 & $ 1.65\pm0.34 $ & $ 1.95\pm0.26$ & [ - ; - ; - ; 185 ; - ] \\
186 & 4\_11 & 18:47:44.78 & -01:54:37.85 & $1.6\times0.95$ & 72 & $ 1.35\pm0.28 $ & $ 1.36\pm0.22$ & [ - ; - ; - ; 186 ; - ] \\
187 & 4\_12 & 18:47:44.79 & -01:54:45.19 & $1.84\times1.57$ & 74 & $ 1.13\pm0.57 $ & $ 1.72\pm0.44$ & [ - ; - ; - ; 187 ; - ] \\
188 & 4\_13 & 18:47:45.29 & -01:54:36.95 & $1.68\times0.92$ & 88 & $ 1.27\pm0.35 $ & $ 0.98\pm0.27$ & [ - ; - ; - ; 188 ; - ] \\
\hline
\multicolumn{4}{l}{Image at 14~kau resolution @85.10~GHz} & & & & \\
189 & 5\_1 & 18:47:47.01 & -01:54:26.56 & $2.8\times2.39$ & 156 & $ 51.92\pm2.9 $ & $ 147.99\pm4.09$ & [ - ; - ; - ; - ; 189 ] \\
190 & 5\_2 & 18:47:46.87 & -01:54:29.56 & $2.93\times2.48$ & 169 & $ 31.15\pm2.21 $ & $ 81.9\pm2.63$ & [ - ; - ; - ; - ; 190 ] \\
191 & 5\_3 & 18:47:46.54 & -01:54:23.19 & $2.81\times2.49$ & 89 & $ 9.31\pm1.61 $ & $ 23.88\pm1.48$ & [ - ; - ; - ; - ; 191 ] \\
&      \\ 
\hline
\end{tabular}
\end{threeparttable}
\end{table*}
}

\end{appendix}

\end{document}